\documentclass[aps, reprint]{revtex4-1} 
\pdfoutput=1
\usepackage{natbib}
\usepackage{bigints}
\usepackage{floatflt}
\usepackage{url}
\usepackage{amsmath}
\usepackage{graphicx}
\usepackage{rotating}
\usepackage{subfigure}
\usepackage{multirow}
\usepackage{latexsym}
\usepackage{textcomp}
\usepackage{verbatim}
\usepackage{color}
\usepackage{pict2e}
\usepackage{subfigure}
\usepackage{everysel}
\usepackage{keyval}
\usepackage{ragged2e}
\usepackage{dsfont}
\usepackage{enumitem}
\usepackage{pstricks}
\usepackage{mathtools}
\usepackage{empheq}
\usepackage{tikz}
\usepackage{ulem}

\usepackage{hyperref}
\hypersetup{
 pdfnewwindow=true, colorlinks=true,
 linkcolor=blue, anchorcolor=blue,
 citecolor=blue, filecolor=blue,
 menucolor=blue, urlcolor=blue}

\begin{document}

\title{Imaging phase slip dynamics in micron-size superconducting rings}
\author{Hryhoriy Polshyn, Tyler R. Naibert}
\affiliation{Department of Physics, University of Illinois at Urbana-Champaign, Urbana, IL 61801, USA}
\author{Raffi Budakian}
\affiliation{Department of Physics, University of Illinois at Urbana-Champaign, Urbana, IL 61801, USA}
\affiliation{Department of Physics, University of Waterloo, Waterloo, ON, Canada, N2L3G1}
\affiliation{Institute for Quantum Computing, University of Waterloo, Waterloo, ON, Canada, N2L3G1}
\affiliation{Perimeter Institute for Theoretical Physics, Waterloo, ON, Canada, N2L2Y5}
\affiliation{Canadian Institute for Advanced Research, Toronto, ON, Canada, M5G1Z8}
\email{rbudakian@uwaterloo.ca}


\begin{abstract}
We present a scanning probe technique for measuring the dynamics of individual fluxoid transitions in multiply connected superconducting structures. 
In these measurements, a small magnetic particle attached to the tip of a silicon cantilever is scanned over a micron-size superconducting ring fabricated from a thin aluminum film. 
We find that near the superconducting transition temperature of the aluminum, the dissipation and frequency of the cantilever changes significantly at particular locations where the tip-induced magnetic flux penetrating the ring causes the two lowest energy fluxoid states to become nearly degenerate.
In this regime, we show that changes in the cantilever frequency and dissipation are well-described by a stochastic resonance (SR) process, wherein small oscillations of the cantilever in the presence of thermally activated phase slips (TAPS) in the ring give rise to a dynamical force that modifies the mechanical properties of the cantilever. 
Using the SR model, we calculate the average fluctuation rate of the TAPS as a function of temperature over a 32-dB range in frequency, and we
compare it to the Langer-Ambegaokar-McCumber-Halperin theory for TAPS in one-dimensional superconducting structures.

\end{abstract}
\thispagestyle{empty}
\thispagestyle{empty}

\maketitle

\section{Introduction}

The single-valuedness of the superconducting wavefunction gives rise to a host of novel macroscopic phenomena, the most striking being fluxoid quantization in multiply-connected devices
  and quantized vortices 
in bulk samples and films~\cite{Tinkham1996}.
The topological nature of fluxoid states makes them robust to small perturbations and endows superconducting rings with the distinct ability to support metastable dissipationless currents.
The behavior of the superconducting phase in  multiply-connected geometries is at the heart of devices of great practical importance, such as  superconducting quantum interference devices (SQUIDs) and flux qubits. New techniques capable of probing and controlling the dynamics of fluxoid states are of great practical and fundamental interest.

A number of experimental  techniques have been applied to study the physics of fluxoid states in superconducting rings, including
transport measurements \cite{Parks1964, Morelle2004}, 
Hall micromagnetometry \cite{Pedersen2001, Vodolazov2003}, 
scanning Hall probe microscopy \cite{Davidovic1996},
SQUID magnetometry \cite{Silver1967, Lukens1968, Zhang1997}, 
scanning SQUID microscopy \cite{Kirtley2003, Bluhm2006, Koshnick2007,Bert2011}, 
calorimetry \cite{Bourgeois2005}, and cantilever torque magnetometry \cite{Petkovic2016, Jang2011}.
Fewer studies have focused on investigating fluxoid dynamics and phase slip rates \cite{Lukens1968, Jackel1974, Zhang1997, Kirtley2003}.
Theoretical studies have addressed fluxoid dynamics as a function of ring geometry and magnetic field \cite{Baelus2000, Berger2003, Kogan2004,Vodolazov2002b}.

Here, we present a scanning probe technique for measuring the dynamics of fluxoid transitions in multiply connected planar superconducting structures. 
In these studies, a micron-size magnetic particle is attached to the tip of an ultra-soft silicon cantilever and scanned over a surface containing an array of lithographically patterned micron-size aluminum rings. During the scan, the cantilever is resonantly driven to a small fixed amplitude using a piezoelectric transducer. 
When the magnetic tip is positioned over an individual ring near the superconducting transition temperature $T_c$, large variations in the frequency and dissipation of the cantilever can be observed at locations where the tip applies a half-integer number of flux quanta through the ring.
The modification to the mechanical properties of the cantilever is caused by the correlated dynamics between the resonant motion of the magnetic tip and thermally-activated phase slips (TAPS) in the ring.
We show that this interaction can be modeled as a classical stochastic resonance (SR) process \cite{Gammaintoni1998}, wherein the frequency and dissipation of the mechanical oscillator are strongly modified when the average fluctuation frequency of TAPS approaches the mechanical resonance frequency of the cantilever. 
A comparison of the relative frequency and dissipation shift provides a direct means of determining the average rate of the TAPS occurring in the ring.

The method introduced in this work is conceptually similar to single-electron electrostatic force microscopy ($e$-EFM) \cite{Woodside2002, Zhu2005, Zhu2008, Bennett2010, Cockins2010, RoyGobeil2015}, in which a similar dynamical effect emerges from the capacitive coupling between the cantilever and a single electron on a quantum dot. In our case, the effect results from the interaction of cantilever with the motion of 'vortices' in a superconducting structure. By analogy, we have termed our technique $\Phi_0$-MFM.

In principle, $\Phi_0$-MFM can be used to study fluxoid dynamics in any multiply connected superconducting structure capable of hosting a discrete spectrum of fluxoid states. 
In this work, we apply $\Phi_0$-MFM to study fluctuations in thin superconducting rings, because the structure of fluxoid states in thin-wall superconducting rings provides a simple framework for demonstrating the concepts behind the technique. 
Furthermore, fluxoid fluctuations in these devices are well-described by the Langer-Ambegaokar-McCumber-Halperin (LAMH) theory for TAPS \cite{Langer1967, McCumber1970}, and they can be compared directly to the experimentally-derived fluctuation rates.

The paper is organized into the following sections:
In section~\ref{sec:Setup}, we discuss the details of the experimental setup. In section~\ref{sec:Results}, we demonstrate the dynamical phenomenon that underlies $\Phi_0$-MFM, and we present a model that considers the dynamics of driven fluxoid transitions and their interaction with the cantilever. 
We use the model to extract the average fluxoid transition rate, and we compare it to the LAMH theory. 
Finally, we present data for a superconducting ring containing a weak link, and we study the phase slips dynamics across the weak link in response to the local fields generated by the magnetic tip. In section~\ref{sec:Conclusion}, we present a summary of the technique and concluding remarks.

\begin{figure*}[htbp]
\includegraphics{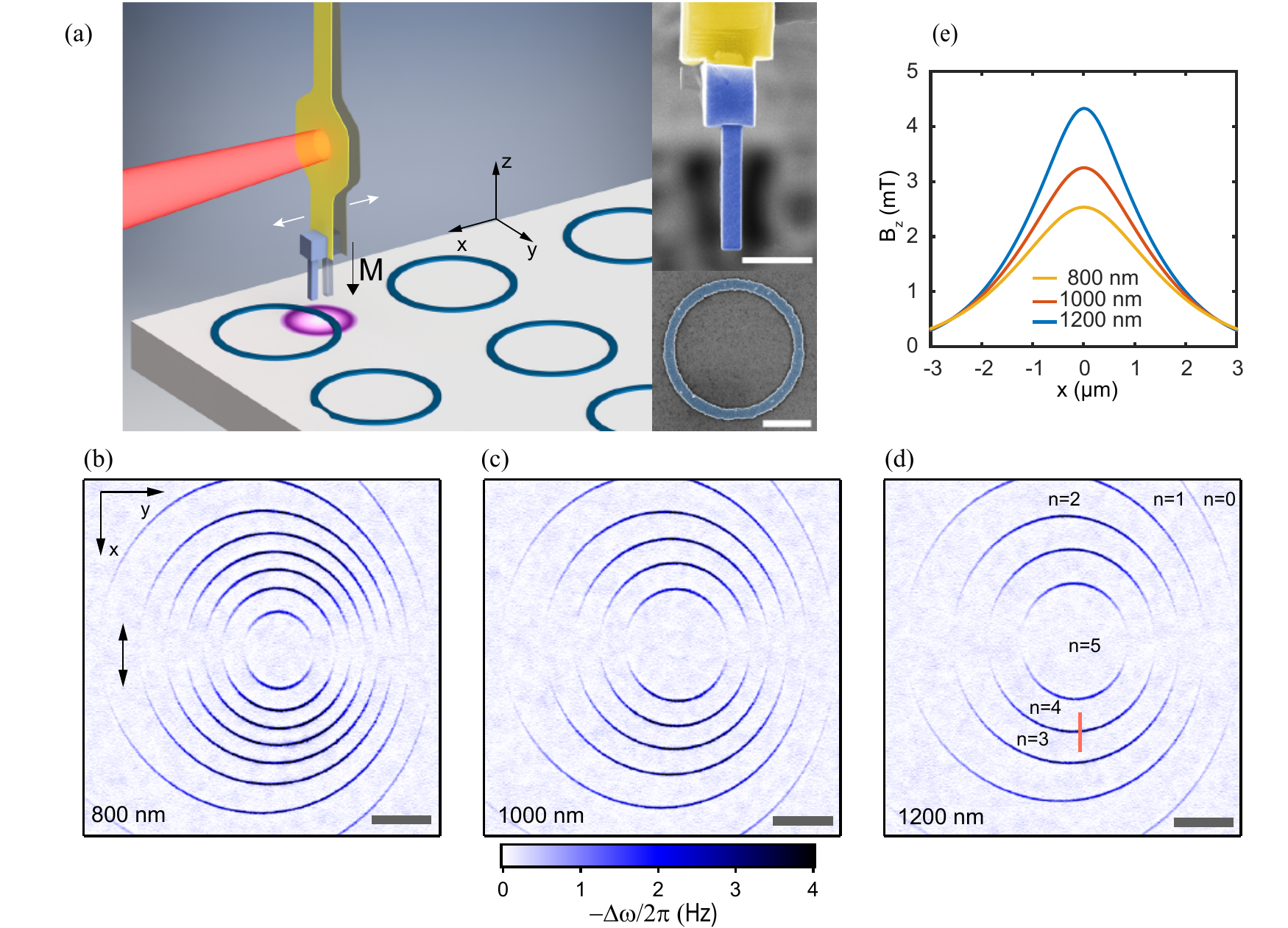}
\caption {(a) Schematic of the experimental setup showing the cantilever positioned over an Al ring. Top inset shows an SEM micrograph of the FIB-shaped $\mathrm{SmCo_5}$ magnetic particle attached to the tip of the cantilever. The magnetic moment of the particle is oriented perpendicular to the surface, in the $z$~direction, and produces a highly inhomogeneous magnetic field in its vicinity (illustrated as the disk shaped region near the tip.) The bottom inset displays an SEM micrograph of Ring~1; (b-d) The $\Phi_0$-MFM images show the frequency shift of the cantilever. 
Dark circular contours correspond to transitions between fluxoid states. 
In the regions between successive transitions, the winding number $n$ of the ring changes by~1.
The images were obtained with a fixed tip-surface separation distance, indicated in the bottom-left-hand corner of each panel. 
The double arrows in (b) indicate the oscillation direction of the cantilever. Changes in the frequency and dissipation of the cantilever across the $n=3$ to $4$ transition, indicated by the red line segment in (d), are presented in detail in Fig.~\ref{fig:StochasticResonance}. (e) Cross section of the magnetic field distribution on the sample surface for various tip-surface heights. The field distributions are estimated from the pattern of fluxoid transitions observed in (b)-(d). All scale bars correspond to~1~$\mu$m.}
\label{Ring}
\end{figure*}

\section{Experimental setup}\label{sec:Setup}
 
The key component of the setup is an ultra-soft silicon cantilever with a magnetic particle attached to the tip (Fig.~\ref{Ring}(a)). 
The cantilever is fabricated from single-crystal silicon with the following dimensions: 80~$\mu\mathrm{m}$ long, 3~$\mu\mathrm{m}$ wide, and 100~nm thick.
The motion of the cantilever is measured by focusing 1510 nm wavelength light from a fiber optic laser interferometer onto the $10\,\mu\mathrm{m}\,\times\,10\,\mu\mathrm{m}$ paddle fabricated near the tip of the cantilever. 
For the measurements presented in Sections~\ref{sec:PlainRing}, the cantilever had a spring constant of $k=1.8 \times  10^{-4}$~N/m, a resonance frequency $\omega_0/2\pi\simeq 7675$~Hz, and a quality factor $Q\simeq 31,800$ at 4~K. 
The measurements in Section~\ref{sec:Constriction}, were performed using a cantilever for which $k=2.3 \times  10^{-4}$~N/m, $\omega_0/2\pi\simeq 7351$~Hz, and $Q\simeq 30,000$.

The cantilever is positioned vertically with respect to the surface in the pendulum geometry, and the tip oscillates in the $x$-direction.
The magnetic tip is fabricated by gluing a micron-size $\mathrm{SmCo_5}$ particle to the tip of the cantilever and shaping it by focused ion beam milling (Fig.~\ref{Ring}(a) (top inset)). 
During the gluing process, an external magnetic field is applied to the $\mathrm{SmCo_5}$ particle to ensure that the magnetic moment of the particle is aligned parallel to the axis of the cantilever.

Arrays of aluminum rings were fabricated by electron-beam lithography and lift-off of 5~nm-thick(45~nm-thick) Ti (Al) films deposited on silicon substrates by electron beam evaporation.
 The substrate containing the patterned devices is mounted onto a three-axis nano-positioner and a scanner that control the relative position of the cantilever with respect to the surface.
The assembly is placed in a high vacuum chamber that is inside of a continuous-flow $^3\text{He}$ refrigerator. 
The sample temperature is controlled using a resistive heater and measured using a calibrated ruthenium oxide thermometer, which are both mounted close to the sample. 
During measurement, the sample temperature can be continuously varied from 340~mK to 4~K with 0.3~mK precision.

We have studied more than 10 rings using 4 different magnetic tips. Here, we report measurements taken on two of these rings.
Ring 1 had a radius of $R=1.40\,\mu \mathrm{m}$ and a uniform wall width of $w=212$~nm (Fig.~\ref{Ring}(a) (bottom inset)).  
Ring 2 had $R=2.38~\mu\textrm{m}$, $w=200$~nm, and a 1.22-$\mu\mathrm{m}$-long constriction, having a minimum width of 60~nm (Fig.~\ref{Constriction1}(a)). The critical temperature and coherence length of these two devices were as follows: Ring 1: $T_c=1.163$~K,  $\xi(0)=108$~nm; and Ring 2: $T_c=1.325$~K, $\xi(0)=104$~nm.
In Appendix~\ref{CohLenAppendix}, we discuss our procedure for determining the $T_c$ and $\xi(0)$
 for the patterned devices.

Force measurements are performed in the frequency detection mode \cite{Albrecht1991}, in which the cantilever is resonantly excited by driving it inside a feedback loop. In our setup, a small piezoelectric transducer is used to apply the feedback signal to the cantilever.
The cantilever frequency is monitored using a phase-locked loop circuit. 
An automatic gain control circuit is used to maintain the desired oscillation amplitude and to monitor the dissipation of the cantilever. 
Images of the cantilever frequency and dissipation are measured by 
exciting the cantilever to a fixed amplitude between 2.5~nm  and 10~nm and scanning it in the $xy$~plane, with the tip positioned at a fixed height above the surface of the sample.

\begin{figure}[htbp]
\includegraphics{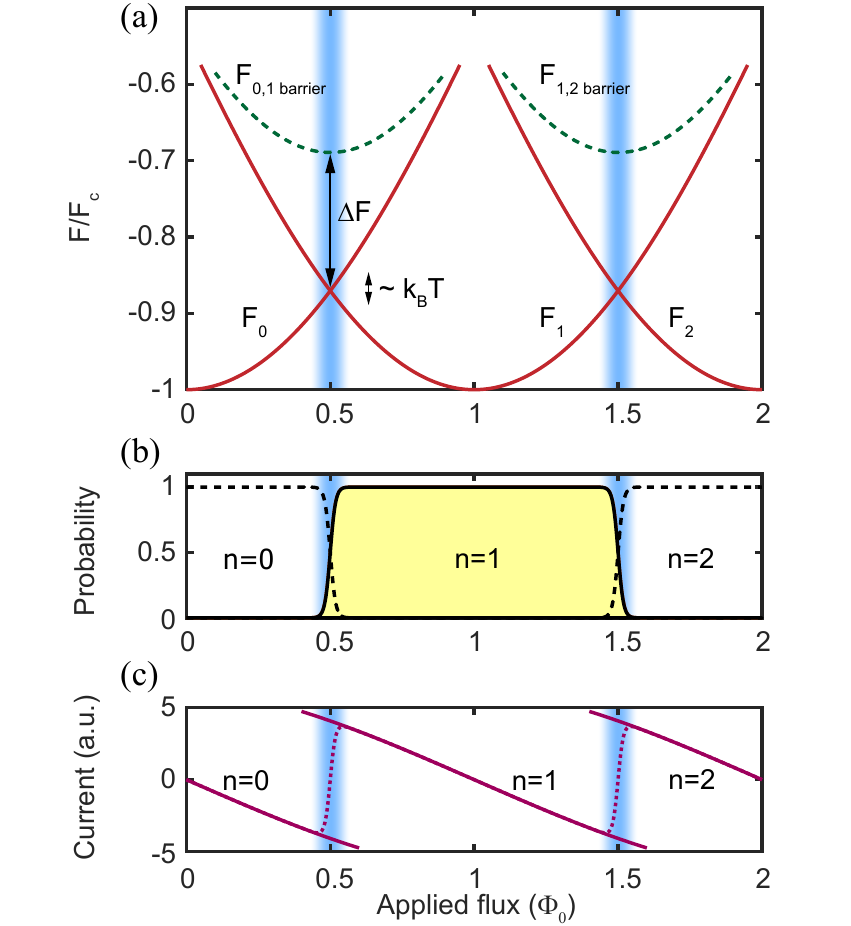}
\caption {Energy, occupation probability, and supercurrent corresponding to the $n=0$ to $2$ fluxoid states. (a) Schematic of fluxoid state energies: solid lines represent energies of the fluxoid states. The dashed lines represent the energy barriers between adjacent fluxoid states. (b) Equilibrium occupation probability for different fluxoid states. (c) The solid lines represent the piecewise-continuous circulating current corresponding to a particular fluxoid state. The dashed line represents the thermal average of the current.
Vertical blue bands on all panels mark the regions, where the energy separation between the states is of the order $\leq k_B T$. 
}
\label{FluxoidStates}
\end{figure}

\section{Results and discussion}\label{sec:Results}
\subsection{$\Phi_0$-MFM imaging of a superconducting ring}\label{sec:PlainRing}

In order for the superconducting order parameter to remain single-valued, the phase of the order parameter must change in integer units of $2\pi$ around any closed path inside the superconductor. 
For a ring geometry, this requirement ensures that the fluxoid, given by $\Phi'=\Phi+(m/e)\oint \mathbf{v}_s \cdot d\mathbf{s} = n\Phi_0$, only takes on integer values $n$ of the flux quantum $\Phi_0 = h/2e$.
Here, $\mathbf{v}_s$ is the superfluid velocity, $\Phi=\oint \mathbf{A} \cdot d\mathbf{s}$ is the total magnetic flux, and $m$, and $e$ are the electron mass and charge, respectively. 
For the present work, the wall thickness of the rings is smaller than both the superconducting penetration depth ($\lambda\sim 1$~$\mu$m) and the coherence length ($\xi \sim 0.5$~$\mu$m). 
Near $T_c$, magnetic screening is negligible and the ring behaves effectively as a one-dimensional (1D) superconductor, with the supercurrent velocity given by $\mathbf{v}_s = \hbar (n-\phi)/2mR$, where $R$ is the radius of the ring, and $\phi=\Phi/\Phi_0$.
By minimizing the Ginzburg-Landau free energy of the ring, we find the free energy of the fluxoid states:
\begin{equation}\label{eq:StateEnergy}
 F_n(\phi)=- F_c  \left(  1 - \frac{ \xi^2}{R^2}\ (\phi-n)^2 \right)^2\,
\end{equation}
where $F_c= V {B_c^2 }/{2\mu_0} $ is the superconducting condensation energy of the ring, $B_c= {\Phi_0}/({2\sqrt{2} \pi \xi\lambda} )$ is the thermodynamic critical field, $V=2 \pi R w d$ is the volume of the ring, and $d$ is the thickness of the film.
The supercurrent in the ring is found from Eq.~\eqref{eq:StateEnergy} using $I=-(1/\Phi_0)\, \partial F/ \partial \phi$:
%
\begin{align}\label{eq:Current}
&I_n(\phi)\simeq-I_0(\phi-n) \left(  1 - \frac{\xi^2}{R^2}\ (\phi-n)^2 \right),\\
&\text{where}\qquad I_0=\frac{\Phi_0}{2 \pi \mu_0 \lambda^2} \frac{wd}{R}.\notag
\end{align}
We note that since for rings in the present work $2 \pi R\gtrsim \xi$, the pair breaking effects are relatively small. Hence $F_n(\phi)$ and $I_n(\phi)$ are close to quadratic and linear functions of the applied flux (Fig.~\ref{FluxoidStates}).

Close to $T_c$, where fluxoid transitions become reversible, the transition between states having winding numbers $n$ and $n+1$ occurs at half-integer values of the flux quantum $\phi=n+1/2$.
For thin-walled superconducting rings, the fluxoid transitions occur via phase slips \cite{Little1967, Langer1967}. 
The metastability of these transitions is related to the height of energy barrier $\Delta F$ connecting two adjacent fluxoid states (Fig.~\ref{FluxoidStates}(a)).
Near $T_c$, the energy barrier becomes sufficiently small and the probability of thermally activated phase slips becomes significant.
In the vicinity of $\phi=n+1/2$, where the separation between adjacent fluxoid states is $|F_{n+1}-F_n|\lesssim k_B T$, thermally activated fluxoid transitions exhibit telegraph-noise behavior \cite{Lukens1968, Kirtley2003}.
At lower temperatures, the height of the energy barrier increases so that $\Delta F \gg k_B T$ and TAPS are exponentially suppressed. 
In this regime, the fluxoid states of the ring exhibit metastability.
Thus, the qualitative behavior of the fluxoid transitions changes from being reversible near $T_c$ to being irreversible and hysteretic at low temperature.

The equilibrium fluxoid state of the ring depends on the applied flux, and hence on the relative position of the magnetic tip and the superconducting ring.
Scans of superconducting rings taken at temperatures sufficiently close to $T_c$ exhibit sharp concentric circular contours in the frequency of the cantilever, corresponding to tip positions where the cantilever frequency dips below the native resonance frequency (Fig.~\ref{Ring} (b-d)).
If the temperature is lowered sufficiently, the states become metastable and the sharp dips in frequency are replaced by much smaller discontinuous jumps. 
The transition between the low- and high-temperature regimes is presented in Appendix~\ref{sec:FTransitions}.
The locations of these features in the images are consistent with tip locations where $\Delta n = 1$. 

In Section \ref{sec:SRmodel}, we show that the frequency dips seen at higher temperatures are caused by a dynamical effect, in which small oscillations of the cantilever, in the presence of TAPS, drive transitions between the two lowest-energy fluxoid states near values of the applied flux that make the energies of the two lowest-energy fluxoid states degenerate.
This effect leads to a synchronization of the fluxoid transitions with the motion of the cantilever (at least in a statistical sense). The resulting interaction of the micromagnet with the synchronously switching supercurrent gives rise to a position-dependent force, which modifies the resonance frequency and dissipation of the cantilever. 
Stationary or quasi-static currents in the ring also produce a frequency shift, however this contribution is often much less than the dynamical one. In particular, near $T_c$, we find that the dynamical contribution ($\Delta f\sim 5$~Hz) is much larger than the static contribution ($\Delta f\sim 0.2$ Hz) and dominates the overall frequency shift.

The dynamical frequency shift maps tip locations to values of the applied flux corresponding to the equilibrium transitions between the lowest-laying fluxoid states. 
For a thin-walled ring, the dark contours seen in the frequency shift image (Fig.~\ref{Ring} (b-d)) correspond to positions where $\phi=n+1/2$.
We note that the dips in frequency are highly spatially localized. 
This feature allows them to be easily distinguished.

Figure  \ref{Ring}(b-d) shows measurements of Ring 1 taken at $T=1.1425$~K for several different tip-surface separations.
The concentric circular patterns observed in these images reflect the fact that the tip-induced magnetic flux through the ring depends primarily on the distance of the tip from the center of the ring.   
The eccentricity of the contours is caused by a slight tilt of the magnetic moment of the $\mathrm{SmCo_5}$ particle with respect to the surface normal (see Appendix~\ref{sec:TipField}). 
We note that the contours begin to fade, and eventually they disappear completely along the line parallel to the $y$ direction 
(horizontal direction in Figs.~\ref{Ring}(b-d))
 and passing through the center of the ring. 
In this region of the scan, $\partial\phi / \partial x = 0$, and small oscillations of the tip in the $x$ direction do not produce a modulation of the magnetic flux.
The regions between the circles correspond to fluxoid states with different winding numbers.
By taking the transition that is farthest from the center of the ring to be the $n=0$ to $1$ transition,
 we can enumerate all of the other observed transitions.
As the tip-surface separation increases, the field on the surface becomes weaker, and fewer transitions are observed.
For a tip-surface separation of 800~nm, the maximum winding number that the tip induces in the ring is $n_{\text{max}}=8$, while for 1000~nm the maximum number is $n_{\text{max}}=6$, and for 1200~nm the maximum number is $n_{\text{max}}=5$.

The spatial map of the fluxoid transitions can be used to estimate the $z$-component of the magnetic field distribution produced by the magnetic tip on the surface. 
To build a model of the magnetic particle, we first measure the total magnetic moment of the particle by cantilever torque
 magnetometry \cite{Stipe2001}, and the dimensions of the particle from the scanning electron microscope (SEM) images of the tip. 
We then calculate an image of the flux generated by the particle through the ring, assuming a uniformly-magnetized tip having the measured dimensions, and we compare it to the data image of the observed transitions for a given tip height. 
To arrive at a more realistic model of the tip, we vary the parameters of the model, including the magnitude, orientation, and distribution of the tip moment, and we match the calculated pattern of fluxoid transitions to those measured from experiment.
The comparison between the calculated frequency shift image and the data is presented and discussed in Appendix~\ref{sec:TipField}.
Estimates of the field profile are shown in Fig.~\ref{Ring}(e). 

\begin{figure}[htbp]
\includegraphics{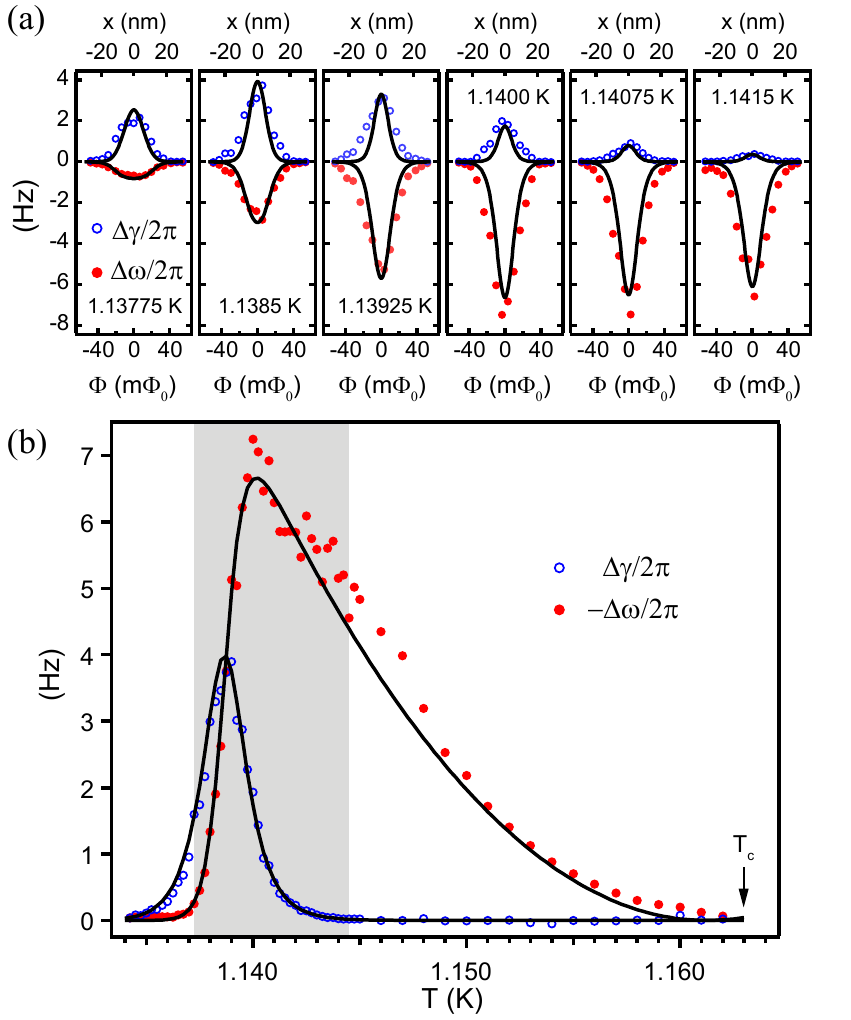}
\caption {Temperature dependence of the signal for the $n=3$ to $4$ transition at the location indicated in Fig.~\ref{Ring}(d).  (a) Line scans across the transition region were obtained at the temperatures indicated in each panel. (b) Plot of the peak frequency (solid circles) and dissipation shift (open circles) as a function of temperature.
Solid lines in (a) and (b) represent the curves calculated using the stochastic resonance model for TAPS (see Section~\ref{sec:PSmeasurements}).
The shaded region in (b) marks the temperature range  for  which the relaxation rate $\nu_r$ was determined.
}
\label{fig:StochasticResonance}
\end{figure}

To study the temperature dependence of the dynamical signal, we took a series of short line scans across the $n=3$ to $4$ transition (marked by a red line segment in Fig.~\ref{Ring}(d)) at different temperatures. 
Figure~\ref{fig:StochasticResonance}(a) shows the cantilever frequency and dissipation shifts for the indicated temperatures. 
The baseline values of the cantilever frequency and dissipation were subtracted from the respective data sets
to isolate the shift caused by the fluxoid dynamics.
To convert the cantilever position to flux (horizontal axis in Fig.~\ref{fig:StochasticResonance}(a)), we obtained an estimate of the conversion factor $\partial\phi / \partial x=1.82 \,\mathrm{\mu m^{ -1}}$ from the spacing of the fluxoid transitions near the region of interest. 
The data plotted in Fig.~\ref{fig:StochasticResonance}(b) represent the peak frequency and dissipation shifts measured from the line scans shown in Fig.~\ref{fig:StochasticResonance}(a) as a function of temperature. 
The line scans were measured using a tip oscillation amplitude of 3.4~nm,  corresponding to a flux modulation amplitude of $6.3\text{m}\Phi_0$. 
The flux modulation amplitude was chosen to be much smaller than the width of the transition region, which, based on the data in Fig.~\ref{fig:StochasticResonance}(a), is about~$36\text{m}\Phi_0$.
The temperature evolution of the frequency and dissipation peak heights is shown in Fig.~\ref{fig:StochasticResonance}(b).
Below 1.135~K, the fluxoid states are metastable and the peaks due to the dynamical effect vanish.
For the range of temperatures between 1.135 and 1.142~K, a rapid increase in dissipation and a decrease in
frequency are observed. 
The height of the dissipation peak reaches its maximum value at 1.1387~K, which is 24~mK below $T_c$ . 
In addition, the height of the dissipation peak decreases and completely disappears by 1.1445 K, while the frequency
peak persists up to $T_c$.

To gain a qualitative understanding of the temperature dependence, observed in Fig.~\ref{fig:StochasticResonance}(b), it is helpful to consider the ratio of the fluxoid transition relaxation rate $\nu_r$ to the cantilever frequency $\omega_0$.
At low temperatures, where $\nu_r/\omega_0\ll1$, the dynamical effect disappears because the height of the energy barrier becomes large, and the tip-induced flux modulation is insufficient to drive fluxoid transitions.
In the high-temperature regime near $T_c$, the energy barrier decreases significantly, so that the equilibrium fluxoid occupation tracks the flux modulation.  
In this regime, $\nu_r/\omega_0\gg1$ and the dynamical force is in phase with the cantilever motion, which shifts the cantilever frequency, but does not change its dissipation.
In the intermediate regime where $\nu_r \sim \omega_0$, a time lag can develop between the equilibrium fluxoid occupation and the cantilever position. The resulting force has components that are in phase and $90^\circ$ out of phase with respect to the cantilever motion, which shifts both the cantilever frequency and dissipation. This dynamical coupling between the cantilever and the fluctuating currents in the superconducting ring can be described in the framework of the SR model \cite{Gammaintoni1998}.

\begin{figure}[h!]
\includegraphics{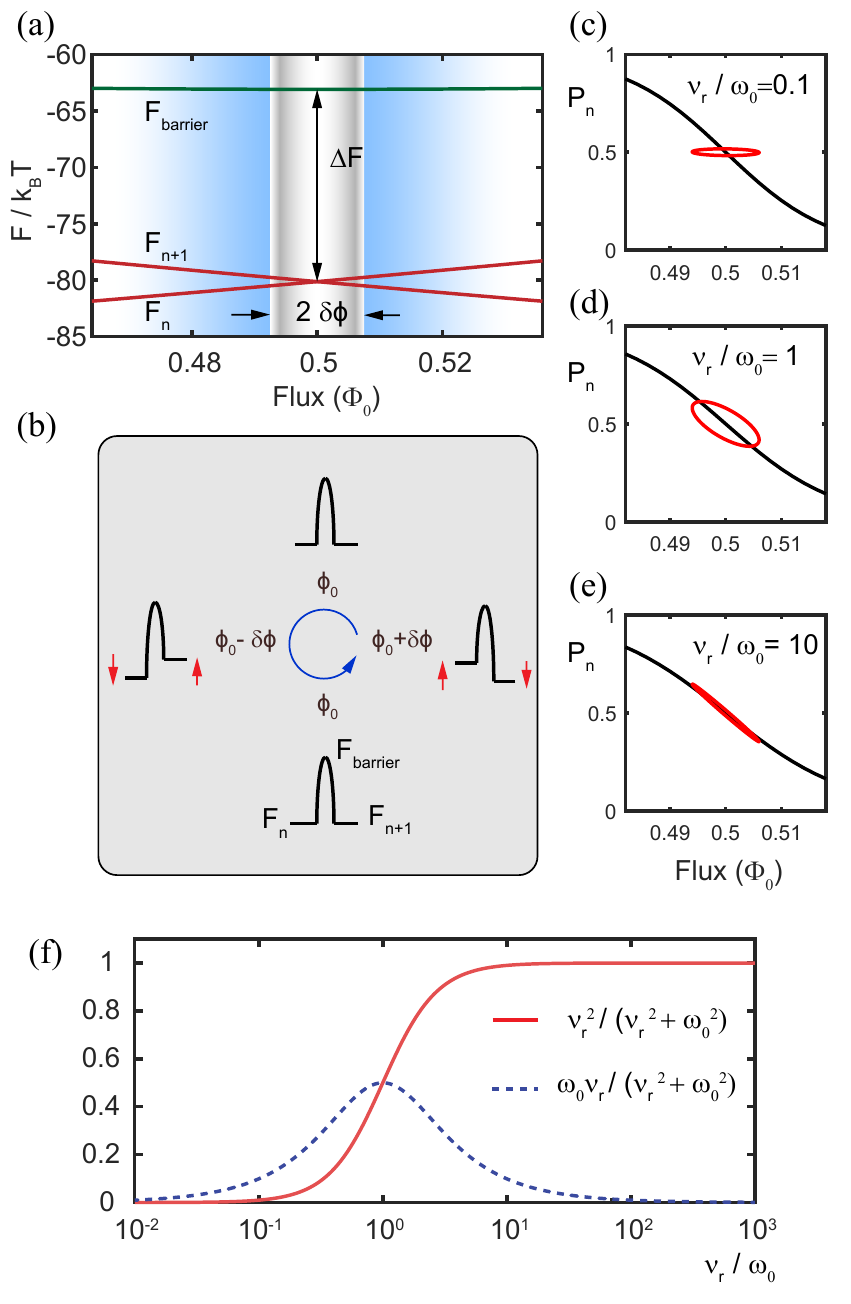}
\caption {Stochastic resonance of TAPSs. (a) Energy diagram showing the region near the $n \leftrightarrow n+1$ fluxoid transition. The width of the gray vertical band represents the extent of the flux modulation due to the cantilever oscillation. 
The area shaded blue represents the range of flux values where the corresponding difference in energy between adjacent fluxoid states is $\lesssim k_BT$.
(b) Schematic diagram showing the modulation of the fluxoid energies caused by the cantilever motion. (c)-(e)
Calculated curves showing the equilibrium $P^{eq}_n$ (solid black) and the non-equilibrium (instantaneous) $P_n$ (red) fluxoid occupation for three different values of $\nu_r/\omega_0$ for one complete cycle of the cantilever motion. (c) The slow relaxation rate prevents $P_n$ from tracking the thermal equilibrium state - weak response. (d) SR condition: synchronization occurs between the fluxoid dynamics and the cantilever motion. The resulting phase lag produces both an in-phase and $90^{\circ}$ out-of-phase response. (e) The fast relaxation rate allows  $P_n$ to track $P^{eq}_n$ - the response is mostly in-phase with the cantilever motion.
(f) Plot of $ \nu_r^2/(\nu_r^2+\omega_0^2)$ (solid) and $\omega_0\, \nu_r/(\nu_r^2+\omega_0^2)$ (dashed) that determine the relative strength of in-phase and out-of-phase components of the response.}
\label{fig:SRscheme}
\end{figure}

\subsection{Cantilever-driven fluxoid transitions in a superconducting ring}\label{sec:SRmodel}

A quantitative description of the experimentally observed dynamical effects requires a model of the coupling between the cantilever motion and the fluxoid dynamics in the superconducting ring. 
For this analysis, it is sufficient to consider a range of energies of order $k_BT$ in the neighborhood of the crossing point between states $n$ and $n+1$, i.e., $|F_{n+1}(\phi)-F_n(\phi))|\lesssim k_B T$, where both fluxoid states have a substantial probability of being occupied (Fig.~\ref{fig:SRscheme}(a)). 
We will assume that the temperature is sufficiently close to $T_c$ so that the energy barrier $\Delta F$ between states $n$ and $n+1$ permits thermally activated transitions, but  far enough from $T_c$ so that the probability of occupying all other states is negligible. 
In this regime, the superconducting fluctuations are governed by the dynamics of the two lowest-energy fluxoid states of the ring.
Thus, the supercurrent $I(t)$ circulating the ring has a two-level stochastic component.
The probability to find the ring in state $n$, when it is in thermal equilibrium, and the cantilever is stationary, is given by
\begin{equation}\label{eq:P_eq}
P^{eq}_n(\phi)=\frac{1}{1+\exp\{-[F_{n+1}(\phi)-F_n(\phi)]/k_B T\}}.
\end{equation}
The dynamics  of the probability $P_n(t)$ is determined by the relaxation rate $\nu_r$:
\begin{align} 
dP_n/dt&=- \nu_r  P_n +\Gamma_{n+1, n}, \label{eq:Pdynamics}\\
\nu_r&=\Gamma_{n, n+1}+\Gamma_{n+1, n}. \label{eq:RelaxRate}
\end{align}
Here, $\Gamma_{n, n+1}$ and $\Gamma_{n+1, n}$ are the transition rates $n\rightarrow n+1$ and $n+1\rightarrow n$, respectively. 
It is worth noting that at $\phi=n+1/2$,  $\Gamma_{n, n+1}=\Gamma_{n+1, n}$  and $\nu_r=2 \Gamma_{n, n+1}$. 

The force produced by the supercurrent $I(t)$ on the cantilever can be expressed as $\zeta (t)=\kappa(\mathbf{r}_{tip}) I(t)$, where $\kappa(\mathbf{r}_{tip})$ represents the coupling, which depends of the relative position of the tip and the ring.
The equation of motion for the cantilever becomes:
\begin{equation}\label{eq:CantileverDynamics}
\ddot x +2\gamma_0 \dot x+\omega_0^2 x = \frac{\omega_0^2}{k} [f(t) +\zeta (t,x)],
\end{equation}
where $x$ is the displacement of the tip from its equilibrium position, $\gamma_0$  is the unmodified dissipation of the cantilever, and $f(t)$ is the force applied by the feedback controller, which resonantly excites the cantilever to a fixed amplitude $x_0$.

The periodic motion of the cantilever tip with amplitude $x_0$ generates a small modulation of the flux through the ring with amplitude $\delta \phi=(d\phi/d x)\, x_0$. 
Small oscillations of flux modulate the energies of the fluxoid states $F_n(\phi)$ and $F_{n+1}(\phi)$, along with the transition rates $\Gamma_{n, n+1}$ and $ \Gamma_{n+1, n}$ as shown in Fig.~\ref{fig:SRscheme}(b). 
In the presence of thermal fluctuations, $\zeta(t)$ and $x(t)$ are statistically correlated. 
The correlation between the force experienced by the cantilever and its position can strongly modify the frequency and dissipation of the oscillator, especially for the case in which the relaxation rate $\nu_r$ matches the cantilever frequency $\omega_0$. 
This phenomenon is generally referred to as stochastic resonance \cite{Gammaintoni1998}.

If $\zeta (t,x)$ is sufficiently small, then the motion of the cantilever can be represented as a sum of two components: $x(t)=x_0 e^{ i\omega t}+x_s(t)$. 
The first term represents the coherent response at the resonance frequency $\omega$ produced by the feedback control, and the second term represents the stochastic part of the motion, with the time-averaged quantity $\langle\hat{x_s} (\omega)\rangle=0$, where $\hat{x_s}(\omega)$ is the Fourier component of the stochastic displacement at the cantilever frequency.
We are interested in the effect of the fluctuating force on the frequency and dissipation of the cantilever. In particular, we consider the time-averaged quantities $\Delta \omega\equiv\left<\omega-\omega_0\right>$ and $\Delta \gamma\equiv \left<\gamma-\gamma_0\right>$,
which are calculated by Fourier transforming Eq.~\eqref{eq:CantileverDynamics}.
\begin{equation}\label{eq:FrequncyDamping}
\Delta \omega \simeq -\frac{\omega_0}{2} \frac{\operatorname{Re}\langle\hat{ \zeta}(\omega)\rangle}{k x_0},\qquad
\Delta \gamma = - \frac{\omega_0}{2} \frac{\operatorname{Im}\langle\hat{\zeta}(\omega)\rangle}{k x_0}.\\
\end{equation}

If the stochastic force due to fluctuating current is weak, i.e., $|\zeta (t,x)|\ll k x_0$, then we expect $|x_s(t)|\ll x_0$, and approximate $x(t)\simeq x_0 e^{ i\omega t}$ in obtaining $\langle\hat{ \zeta}(\omega)\rangle$.
This  approximation allows us to effectively decouple the cantilever dynamics from the dynamics of the phase slips, which greatly simplifies the analysis.

If the flux modulation is sufficiently small such that $(dP^{eq}_n/d\phi) \delta \phi   \ll 1$, the resulting modulation of $P_n(t)$ is linear in $\delta \phi$, with
\begin{align}\label{eq:Psolution}
&P_n(t)\simeq P_n^{eq}(\phi_0) +\delta P e^{i(\omega t-\theta)},\\
&\delta P=\frac{dP^{eq}_n}{d\phi} \delta \phi\cos{\theta}, \label{eq:PsolutionAmp}\\
&\theta=\arctan\left(\frac{\omega}{\nu_r}\right).\label{eq:PsolutionPhase}
\end{align}
Figs.~\ref{fig:SRscheme}(c-e) shows the dynamics of $P_n(t)$ for one complete cycle of the cantilever motion,   calculated using  Eqs.~(\ref{eq:Psolution} -\ref{eq:PsolutionPhase}) for three different values of $\nu_r / \omega_0$. 
%

To find $\Delta \omega$ and $\Delta \gamma$ we consider the ensemble averaged current:
$\langle{ I(t)}\rangle=I_n(\phi) P_n(t)+I_{n+1}(\phi) (1- P_n(t))$.
Using Eq.~\ref{eq:Current}, we find that  
 $\langle{ I(t)}\rangle=-\Delta I P_n(t)-I_{n+1}(\phi(t))$, where $\Delta I (\phi) =I_{n+1}(\phi)-I_{n}(\phi)$.
The first term in $\langle I(t) \rangle$ describes the contribution to the current from the thermally-activated transitions between the two states, and the second term represents the flux dependence of the current in each state.
Here, we consider only the first term, since the second term is not relevant to the effect of interest. 
The resulting expression for the average stochastic force is $\langle\zeta (t)\rangle=-\kappa(\mathbf{r}_{tip}) \Delta I P_n(t)$, and the Fourier component of the statistically-synchronized stochastic force due to the cantilever-driven phase slips is $\langle\hat{ \zeta}(\omega)\rangle=-\kappa(\mathbf{r}_{tip}) \Delta I \delta P e^{-i \theta}$.
By combining this expression with Eq.~\eqref{eq:FrequncyDamping}, we find the following expressions for the changes in frequency and dissipation
\begin{align}
&\Delta \omega \simeq \frac{\omega_0}{2}\, \alpha\, \beta\, \frac{ \nu_r^2}{\nu_r^2+\omega_0^2}, 
\label{eq:Frequency}\\
&\Delta \gamma = -\frac{\omega_0}{2}\,\alpha \, \beta \,\frac{\omega_0\, \nu_r}{\nu_r^2+\omega_0^2}, 
\label{eq:Damping}
\end{align}
where
\begin{equation}
\alpha=\frac{\kappa(\mathbf{r}_{tip})  }{k} \frac{d \phi}{dx},\qquad
\beta=\Delta I (\phi) \frac{dP^{eq}_{n}}{d\phi}. 
\label{eq:alpha}
\end{equation}

Notice that $\alpha$ depends only on parameters of cantilever and its position with respect to the ring.
We find that the dissipation shift vanishes for both $\nu_r\ll \omega_0$ and $\nu_r\gg \omega_0$, and it becomes maximum for $ \nu_r\sim \omega_0$ (Fig.~\ref{fig:SRscheme}(f)). 
On the other hand, the frequency shift is small for $ \nu_r\ll \omega_0$, increases rapidly with  $\nu_r$ for $ \nu_r\sim \omega_0$, and gradually decreases when $ \nu_r> \omega_0$ because of the temperature dependence of  $\beta$ (see Eq.~\eqref{eq:BetaCrossing}). 
The width of the frequency dips and dissipation peaks depends on the range of flux values for which $dP^{eq}/d\phi$ is sufficiently large.

By constructing the ratio $\Delta\omega/\Delta\gamma$ using Eqs.~\eqref{eq:Frequency} and \eqref{eq:Damping}, we find a simple expression that involves only the average fluctuation rate in terms of the cantilever frequency:
\begin{equation}\label{eq:phaseSlipRate}
\frac{ \nu_r}{\omega_0}=\frac{ \Delta \omega}{\Delta \gamma}.
\end{equation}
It is worth emphasizing that, while the expressions for $\Delta\omega$ and $\Delta\gamma$ are each functions of position and temperature, the ratio  $\Delta\omega /\Delta \gamma$  in the linear SR regime only depends on the phase slip rate. 
Thus, the ratio provides a robust and convenient way to measure $\nu_r$ without any prior knowledge of $\alpha$ and $\beta$.

\subsection{Phase slip rate measurements} \label{sec:PSmeasurements}

\begin{figure}[htbp]
\includegraphics{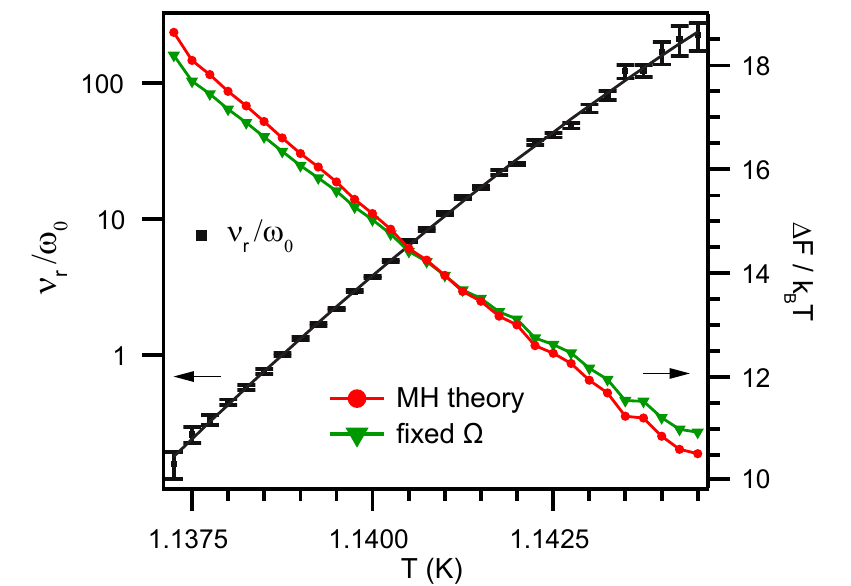}
\caption {Left axis: $\nu_r/\omega_0$ determined from the data shown in Fig.~\ref{fig:StochasticResonance}(b). Right axis: Phase slip energy barrier $\Delta F$ derived from $\nu_r$. Solid circles correspond to $\Delta F$ derived using McCumber-Halperin expression for attempt frequency $\Omega$, solid triangles correspond to constant $\Omega=3\times 10^{11}$~$\text{s}^{-1}$.
}
\label{fig:SRdataAnalysis}
\end{figure}

To be in the linear SR regime, two conditions must be satisfied: (i) the oscillation amplitude of the cantilever should be sufficiently small i.e., $(dP_n^{eq}/d\phi)  (d\phi/dx)  x_0 \ll 1$, so that the modulation of the occupation probability of the fluxoid states is small; (ii) the stochastic force acting on the cantilever should be a weak perturbation, i.e. $|\zeta(t,x)|\ll k x_0$.  
For data shown in Fig.~\ref{fig:StochasticResonance}, the first condition is satisfied since the oscillation amplitude of the cantilever is much smaller than the observed width of the frequency and dissipation peaks. 
We assume that the second condition is met since the observed frequency shifts do not exceed 1\% of the native resonant frequency of the cantilever.
Near $\phi = n+1/2$ and using Eq.~\eqref{eq:phaseSlipRate}, we can directly determine the fluxoid transition rate $\nu_r$.
Figure~\ref{fig:SRdataAnalysis} shows a plot of $\nu_r$  calculated using the data between 1.1372~K and 1.1445~K (shaded region in Fig.~\ref{fig:StochasticResonance}(b)) and  Eq.~\eqref{eq:phaseSlipRate}.
For this range of temperatures, $\nu_r$ increases nearly exponentially from  $0.16\omega_0$ to $224\omega_0$, or from  $\left. 7.7\times 10^{3}\, \text{s}^{-1}\right.$ to $\left. 10.8 \times 10^{6}\, \text{s}^{-1}\right.$, and it can be approximated by 
\begin{equation}
\ln(\nu_r/\omega_0)=\tilde{t}\,1107-\tilde{t}^2\,27\times 10^{3},
\label{eq:ApproxRate}
\end{equation}
where $\tilde{t}=T- 1.13875$~K.

The height of the phase slip energy barrier $\Delta F$ can be determined from  the measured $\nu_r$.
The rate of TAPS is given by  
\begin{equation}
\Gamma(T)=\Omega \exp (-\Delta F/k_B T) \label{eq:PLrate},
\end{equation}
where $\Omega$ is an attempt frequency.
At $\phi=n+1/2$, $\Gamma_{n, n+1}=\Gamma_{n+1, n}=\Gamma$  and $\nu_r=2 \Gamma$.
We use the result obtained by McCumber and Halperin~(MH)~\cite{McCumber1970} to determine the attempt frequency $\Omega$:
\begin{equation}
\Omega = (2\pi R/\xi)(\Delta F/k_BT)^{0.5}/\tau ,\label{eq:PLrateAF}
\end{equation}
where $\tau=\pi \hbar / 8 k_B (T_c-T)$. 
Using Eqs.~(\ref{eq:PLrate}-\ref{eq:PLrateAF}) and  the temperature dependence for the coherence length $\xi=\xi(0)(1-T/T_c)^{-1/2}$ (for the measurement of $\xi(0)$ see Appendix~\ref{CohLenAppendix}), 
we calculate  the temperature dependence of $\Delta F$ (Fig.~\ref{fig:SRdataAnalysis}).
While Eq.~\eqref{eq:PLrateAF} is crucial for calculating the absolute value of $\Omega$, its effect on the temperature dependence of $\Gamma (T)$ is small. 
For the temperature range shown in Fig.~\ref{fig:SRdataAnalysis}, $\Omega$ changes only from $4.5\times10^{11}$~$\text{s}^{-1}$ to $2\times10^{11}$~$\text{s}^{-1}$, while $\nu_r$ changes by three orders of magnitude.
 If the value of $\Delta F$ is calculated 
 assuming a constant value  of $\Omega=3\times10^{11}$~$\text{s}^{-1}$ in this temperature range,  the resulting $\Delta F$ deviates by less than 3\% from the value calculated from MH theory (Eq.~\eqref{eq:PLrateAF}), as shown in Fig.~\ref{fig:SRdataAnalysis}. 

As described in details in Append.~\ref{sec:TipRingCoupling}, from the SR model (Eqs.~(\ref{eq:Frequency}-\ref{eq:Damping})) together with the measured $\nu_r(T)$ (Eq.~\eqref{eq:ApproxRate}) and the estimate of $\alpha$,  we find $\beta(T)$ and also an estimate for $\lambda(T)= 164/\sqrt{(1-t)}$~nm.
The estimated values of $\lambda(0)$  and $\alpha$, together with  the measured $\nu_r(T)$, allow us to reproduce both the 
temperature dependence of $\Delta \omega(T)$ and $\Delta \gamma(T)$ at $\phi=n+1/2$ (Fig.~\ref{fig:StochasticResonance}(b)), as well as the shape of $\Delta \omega(\phi)$ and $\Delta \gamma(\phi)$ peaks as a function of the magnetic field (Fig.~\ref{fig:StochasticResonance}(a)) (see Append.~\ref{sec:TipRingCoupling}).

It is instructive to compare the determined phase slip barrier height $\Delta F$ to the value, predicted by the theory developed by Langer and Ambegaokar~(LA).
The energy barrier for phase slips in a 1D wire is  $\Delta F =(8\sqrt 2/3) \xi wd\, {B_c^2 }/{2\mu_0}$~\cite{Langer1967}.
This result has been generalized for thin-walled superconducting rings~\cite{Zhang1997}.
For rings whose circumference is large with respect to the coherence length ($\xi/2\pi R\approx0.09$ for Ring~1), the saddle point free energy $F_{barrier}$ for the $n\rightarrow n+1$  transition near $\phi\approx n+0.5$  is~\cite{Zhang1997, Bert2011}:
\begin{equation}
F_{barrier}\simeq -F_c+ \frac{8\sqrt 2}{3} \frac{B_c^2 }{2\mu_0} \xi wd,       
\label{eq:SaddlePointEnergy}
\end{equation}
and the corresponding barrier height is 
\begin{multline} \label{eq:BarrierEnergy}
 \Delta F= F_{barrier}-F_n(\phi)\simeq \frac{8\sqrt 2}{3} \frac{B_c^2 }{2\mu_0} \xi wd \,-\\
- F_c \left[\frac{2 \xi^2}{R^2}(\phi-n)^2-\frac{\xi^4}{R^4}(\phi-n)^4\right].
\end{multline}
We find that $\Delta F$ calculated from Eq.~\eqref{eq:BarrierEnergy} using a previously determined value of penetration depth $\lambda (0)\simeq 164$~nm (see Append.~\ref{sec:TipRingCoupling}) is significantly higher than the value determined from the measured $\nu_r$ and MH theory. 
For example, at 1.141~K, the value of  $\Delta F$  predicted using LA theory is $24.2k_{B}T$, in contrast to $14k_{B}T$ derived from $\nu_r$~(see Fig.~\ref{fig:SRdataAnalysis}).  
%
We note that this change in $\Delta F$ corresponds to only an 11\% change in the saddle-point energy $F_{barrier}$ with respect to $F_c$.(Eq.~\ref{eq:SaddlePointEnergy}).
There could be two reasons that cause the phase slip energy barrier to be lower than the value predicted by Eq.~\eqref{eq:BarrierEnergy}.
First, the film surface roughness could lower the barrier height.
Second, the magnetic field produced  by the tip might decrease the phase slip barrier.
The peak value of the magnetic field under the tip is about 2.5~mT, which is comparable to the 8~mT homogeneous magnetic field that destroys  the superconductivity in the ring at 1.14~K (see Fig.~\ref{fig:Tc}(b) and discussion in Append.~\ref{CohLenAppendix}).
However, a detailed theoretical analysis of this effect is beyond the scope of this paper.

\begin{figure}[h!]
\includegraphics{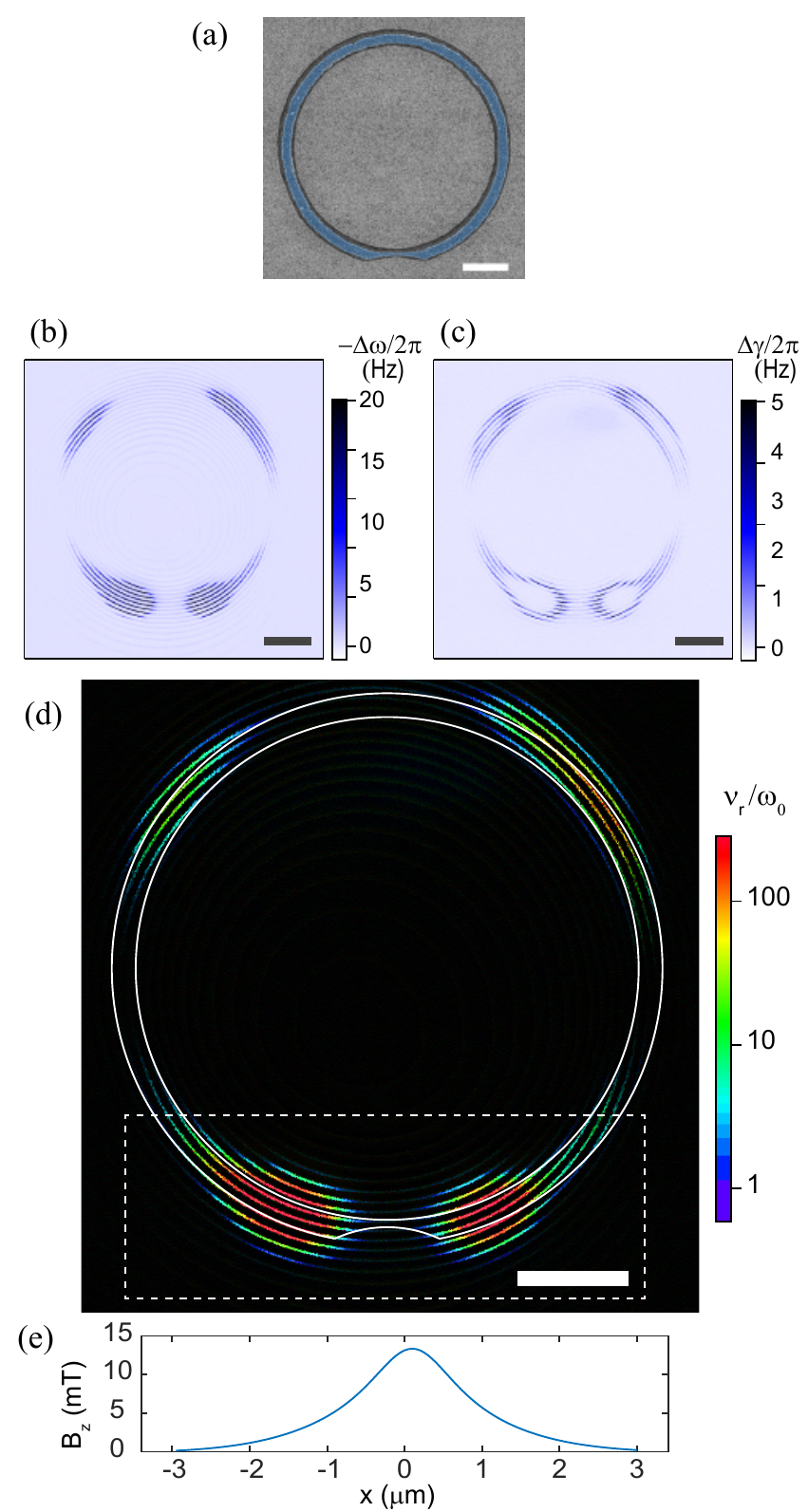}
\caption {(Color online) Stochastic resonance imaging of a thin ring containing a constriction.
The stripes in the images correspond to individual fluxoid transitions. (a) An SEM image of Ring 2. The ring has a radius
$R=2.38~\mu\textrm{m}$, a width $w=200$~nm and a 1.22-$\mu\mathrm{m}$-long constriction, having a minimum width of 60~nm.
(b) Frequency shift $\Delta \omega$. (c) Dissipation shift $\Delta \gamma$ (d) Phase slip rate $\nu_r$. The solid white lines indicate the outline of Ring 2. 
The dashed line marks the rectangular region shown in Fig.~\ref{Constriction2}.
 (e) Cross section of the magnetic field distribution on the surface.
Measurements were performed at $1.280$~K for a tip-surface separation of 650~nm. All scale bars are 1 $\mu$m .}
\label{Constriction1}
\end{figure}

\begin{figure}[htbp]
\includegraphics{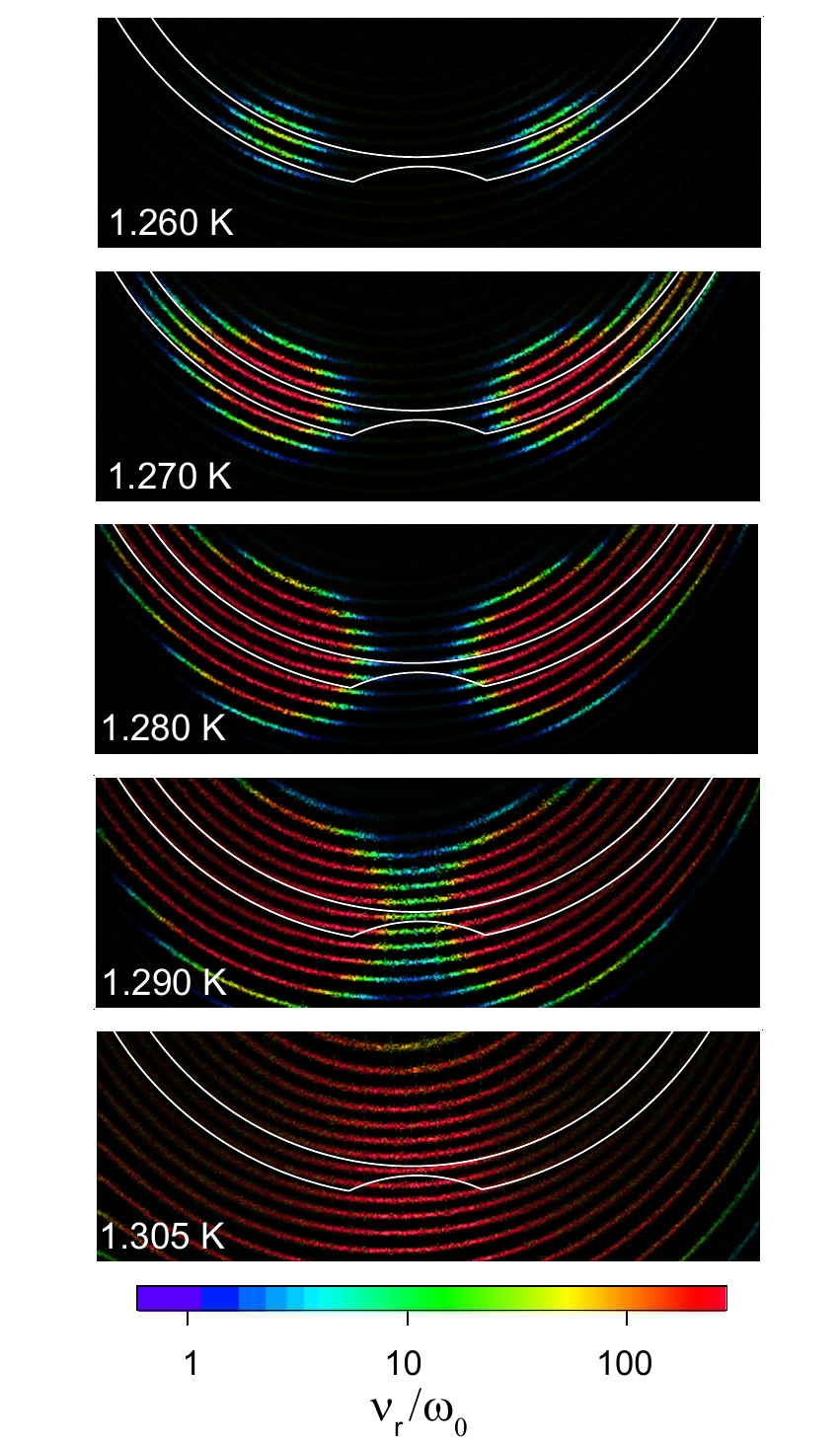}
\caption {(Color online) Temperature evolution of $\nu_r$ taken for the rectangular region indicated in Fig.~\ref{Constriction1}(d). Measurements were performed at a tip-surface separation of 550~nm.}
\label{Constriction2}
\end{figure}

\subsection{Stochastic resonance imaging of the phase slip rate in a ring containing a constriction}\label{sec:Constriction}
In this section, we demonstrate the ability to use the strong magnetic fields produced near the magnetic tip to locally probe the field dependence of the phase slip rate in a thin ring containing a constriction. By combining the frequency and dissipation shift images using Eq.~(\ref{eq:phaseSlipRate}), we construct an image that shows the phase slip rate as a function of tip position. 
As an example, we present a qualitative study of an aluminum ring containing a  constriction. 
An SEM image of the device is shown in Fig.~\ref{Constriction1}(a).

We use the frequency and dissipation shift images (Fig.~\ref{Constriction1}(b-c)) to construct an image of the phase slip rate shown in Fig.~\ref{Constriction1}(d). 
The color in Fig.~\ref{Constriction1}(d) represents the quantity $\Delta \omega/\Delta \gamma$, and the brightness represents the magnitude of the signal, $\sqrt{\Delta \omega^2+\Delta \gamma^2}$. 
This representation is chosen to emphasize only those parts of the image for which $\Delta\omega$ or $\Delta\gamma$ is sufficiently large, so as to minimize the error in the ratio $\Delta\omega /\Delta \gamma$.
Red  and blue colors correspond to tip positions for which $\nu_r > \omega_0$ and $\nu_r < \omega_0$, respectively.

For these measurements, we needed the magnetic field generated by the tip to be large enough to locally suppress the superfluid density in the aluminum ring. To achieve the necessary field, we attached a larger magnetic particle to the cantilever and positioned the tip closer to the surface. 
A cross section of the estimated tip field profile is shown in figure~\ref{Constriction1}(e).
The peak magnetic field under the tip for this tip is $\sim 13$~mT ($\sim 5\times$ larger than the tip field realized for the measurements on Ring 1).
The full width at half-height of the field profile is $\sim 1.6\, \mu m$.

From the image in Fig.~\ref{Constriction1}(d)  we observe that the phase slip rate is the lowest when the tip is positioned directly above the constriction, higher when it is located over the ring far away from the constriction, and the highest when it is positioned on the wider portion of the ring immediately adjacent to the constriction. Figure~\ref{Constriction2} shows the temperature evolution of the phase slip rate along the constriction. We observe that as the temperature is increased, the regions next to the constriction are the first to undergo SR at $\sim 1.26$~K, followed by the portion of the ring away from the constriction at $\sim 1.27$~K, and lastly the constriction itself at $\sim 1.29$~K.
This finding indicates that the tip field lowers the energy barrier most effectively when the tip is positioned on either side of the constriction, but not directly over it. This somewhat counterintuitive finding highlights the unique capability of $\Phi_0$-MFM to use the strong magnetic fields produced by the tip to study the local properties of a micron-scale superconducting device.
The effect is quite robust; similar behavior was observed on four different structures containing constrictions using three different magnetic tips.

While a quantitative explanation of these observations requires a numerical simulation of the Ginzburg-Landau equations and goes beyond the scope of this paper, the observed field dependence of the phase slip rate can be qualitatively understood from the following considerations.
In the case of a homogeneous wire, the energy barrier for a phase slip is of the order of the energy needed to suppress the order parameter in a length $\sim \xi$ of the wire.
When the magnetic tip is placed above the wire, the magnetic field induces a whirlpool of current in the superconducting region below the tip, which locally suppresses the order parameter, and consequently lowers the energy barrier locally near the tip.
The energy barrier for a superconducting ring of variable cross section placed in an inhomogeneous magnetic field is determined by its weakest part, where a combination of the sample geometry and the magnitude of the order parameter minimizes the energy barrier. 

We find that the largest suppression of the energy barrier is achieved when the tip is located adjacent to, but not directly above, the constriction. While our measurements cannot determine the exact location where the phase slip occurs, the following scenario could explain the observed behavior. Superconductivity in the wider section of the wire is suppressed more strongly by the tip field because the critical field for a wire of width $w<\lambda$ scales inversely with the width (see \eqref{eq:Bc||}). Therefore, there is a greater suppression of the order parameter when the tip is positioned in the regions adjacent to the constriction, rather than directly over the constriction. The suppressed order parameter propagates into the constriction over a distance$~\sim\xi$ via a negative proximity effect. Here, $\xi(1.28 ~K)\simeq 560$~nm, is comparable to the length of the constriction. The reduction of the order parameter together with the smaller cross-sectional area of the constriction lowers the barrier in this region further, thus increasing the phase slip rate through the constriction. 
 
\section{Conclusion}\label{sec:Conclusion}

We have introduced a scanning probe technique, $\Phi_0$-MFM, for studying phase slip dynamics in multiply connected superconducting structures. 
In $\Phi_0$-MFM, the dynamical interaction between a magnetic particle attached to the cantilever and the fluctuating currents in a superconducting device modifies the frequency and dissipation of the cantilever. We find that over a wide range of fluctuation frequencies, the interaction is well described by a linear SR process. We further demonstrate that the SR model can be used to extract the average rate of TAPS in thin-wall superconducting rings. 
We find that the measured phase slip rate is consistent with thermally activated behavior, but the corresponding energy barrier is reduced in comparison to the Langer-Ambegaokar prediction.
Lastly, we use a superconducting ring containing a constriction to demonstrate that the strong magnetic field produced by the magnetic particle may be used to probe the effects of a local magnetic field on the energy barrier of the fluxoid states.

In summary, $\Phi_0$-MFM is a non-contact scanning probe technique capable of mapping out fluxoid or vortex transitions and characterizing their dynamics over a wide range of temperatures and magnetic fields. This technique could be a valuable tool for investigating various superconducting structures, with applications to fundamental science and technology.

\section{Acknowledgments}

 We are grateful to Nadya Mason for useful discussions. This work was supported by the DOE Basic Energy Sciences under Contract No.~DE-SC0012649, the Department of Physics and the Frederick Seitz Materials Research Laboratory Central Facilities at the University of Illinois.

\appendix 
\section{Critical temperature and coherence length measurements}\label{CohLenAppendix}

\begin{figure}[h!]
\includegraphics{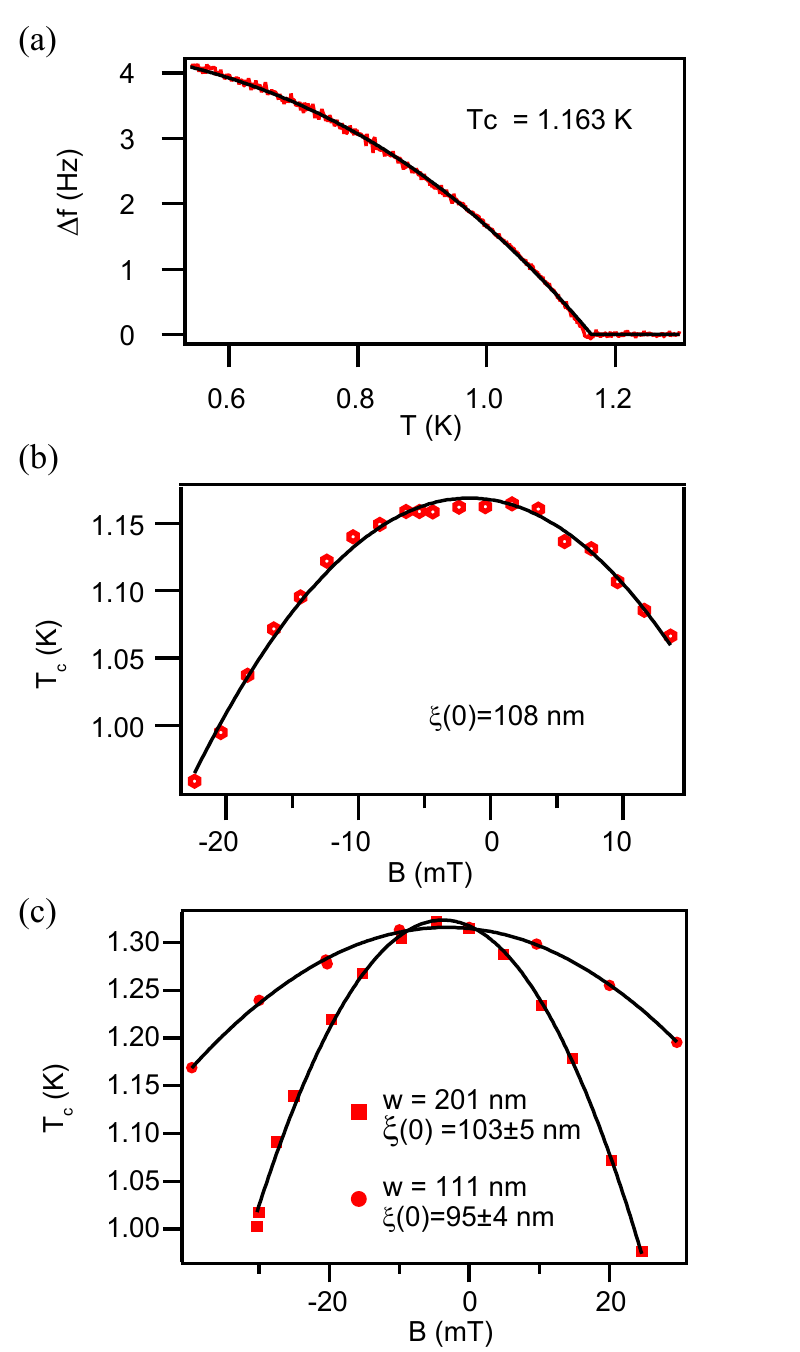}
\caption {Critical temperature and coherence length measurements. (a) Frequency shift as a function of temperature for Ring~1. The black line is a fit to the function $\Delta f=\Delta f_0[1-(T/T_c)^3]$. (b) $T_c$ as a function of magnetic field applied perpendicular to the plane of Ring~1. The solid line represents a fit to Eq.~\eqref{eq:TcH}. 
(c) $T_c$ vs. B measurements for two rings with the same radius $R=0.9\mu$m but different wall width $w=201\pm8$~nm and $w=111\pm5$~nm. The solid lines represent  fits to Eq.~\eqref{eq:TcH}. }
\label{fig:Tc}
\end{figure}

The critical temperature of the aluminum rings is determined by monitoring the resonant frequency shift as a function of temperature. For these measurements, the magnetic particle is placed at a fixed location above the wall of the ring.
Data obtained for Ring 1  are shown in Fig.~\ref{fig:Tc}~(a).
We confirmed that the tip location did not significantly affect the value of $T_c$ by varying the tip-surface height.

The frequency shift is expected to be proportional to the supercurrent in the ring.
We found that the temperature dependence  $\Delta f(T) \propto 1-(T/T_c)^3$ provides excellent agreement with the observed temperature dependence of the frequency shift. 
Based on the onset of the frequency shift, we determine the critical temperatures to be $T_c$=1.163~K (Ring 1) and $T_c$=1.325~K (Ring 2).
Ring~1 and Ring~2 were fabricated separately using two different evaporators for Al film deposition. 
This explains the significant difference between critical temperatures of the two rings.

The superconducting coherence length was determined from the suppression  of $T_c$ with the magnetic field applied perpendicular to the plane of the ring.
The magnetic field is generated using a  superconducting solenoid magnet. 
The critical field of a thin-wall ring, for magnetic fields applied perpendicular to the plane of the ring, is analogous to the parallel critical field of a thin film, provided that $w\ll\lambda$ \textcolor[rgb]{0,0,1}(see p.~131 of \cite{Tinkham1996}):
\begin{equation}\label{eq:Bc||}
B_{c\Vert}=2\sqrt{6} B_c\frac{\lambda}{w}.
\end{equation}
Equation~\eqref{eq:Bc||} holds for thin superconducting rings, because the demagnetization effect vanishes at the second order superconducting transition where $\lambda \to \infty$.
By substituting $B_c= {\Phi_0}/({2\sqrt{2} \pi \xi\lambda} )$  and  $\xi(T)=\xi(0)/\sqrt{1-T/T_c}$  we find:
\begin{equation}
B_{c\Vert} (T)=B_{c\Vert} (0) \sqrt{1-T/T_c},
\end{equation}
where$\quad B_{c\Vert} (0)=\frac {\sqrt{3}}{\pi} \frac {\Phi_0}{w \xi(0)}$. Thus, the superconducting transition temperature is a quadratic function of applied field.
\begin{equation}\label{eq:TcH}
T_c(B)=T_c(0) \left[1-\frac{B ^2} {B_{c\Vert}^2 (0)} \right]
\end{equation}
This behavior is indeed observed in measurements (Fig.~\ref{fig:Tc}~(b)). 
The value of $B_{c\Vert} (0)$ is found by fitting to Eq.~\eqref{eq:TcH}. 
Based on this fit, the value of the superconducting coherence length is
\begin{equation}\label{eq:xi}
\xi(0)= \sqrt{3} \Phi_0/(\pi w B_{c\Vert} (0)). 
\end{equation}
For  Ring~1, we find $\xi(0) =108 \, \text{nm}$. 
Similar measurements of the coherence length gave $\xi(0) =104 \, \text{nm}$ for Ring~2.

As a control experiment for using Eqs.~(\ref{eq:TcH}-\ref{eq:xi}), we measured the suppression of the transition temperature by a magnetic field in two rings of the same radius $R=0.9\mu$m, but different wall widths $w=201\pm8$~nm and $w=111\pm5$~nm (Fig.~\ref{fig:Tc}~(c)). 
These two rings were close to each other on the same chip.
The ring with wider walls shows higher critical fields, which is consistent with Eq.~\eqref{eq:Bc||}. 
The values of coherence length $\xi(0)$, derived from Eqs.~(\ref{eq:TcH}-\ref{eq:xi}), are  $103\pm5 \, \text{nm}$ and 
$95\pm 4 \, \text{nm}$ for rings with wider and narrower walls respectively.
The two values are in reasonable agreement with each other. 

\section{Fluxoid transitions at lower temperatures}\label{sec:FTransitions}

\begin{figure}[h!]
\includegraphics{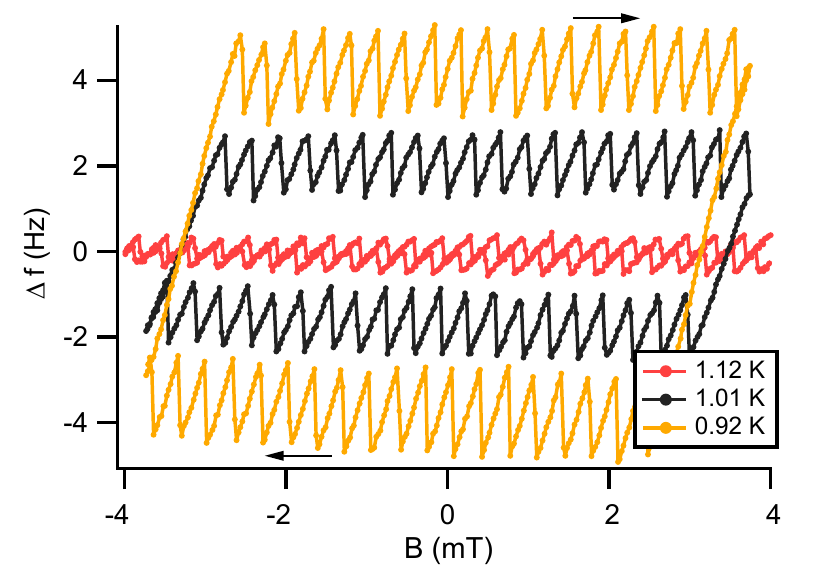}
\caption {Field sweeps obtained for Ring 1, with the magnetic tip positioned 600~nm above the center of the ring. The arrows indicate the direction of the field sweep.}
\label{fig:FSweeps}
\end{figure}

The field sweep curves obtained at lower temperatures reveal fluxoid transitions with a period consistent with fluxoid quantization (Fig.~\ref{fig:FSweeps}).
With the cantilever positioned 600~nm above the center of Ring 1, the shift in the resonant frequency of the cantilever was recorded as a function of the external magnetic field, applied using the superconducting magnet. The data were obtained by cooling the sample in zero field, and sweeping the direction of the magnetic field in a closed cycle, indicated by the arrows in Fig.~\ref{fig:FSweeps}. 
The jumps in frequency correspond to individual fluxoid transitions, with a period of $0.339\pm 0.001$~mT.
The measured period is in excellent agreement with the calculated value of 0.336~mT for a ring of radius R=1.4~$\mu$m.

The large hysteresis observed at low temperature is a consequence of the increased barrier height of the fluxoid transitions, which prevents the small variations of the magnetic flux caused by the cantilever oscillations from changing the fluxoid state of the ring. 
Thus, the discrete frequency jumps observed at low temperature originate from an interaction that is different from the dynamical one discussed in the main paper. 
At lower temperature, the current in the ring is independent of the cantilever oscillation, and the frequency shift, caused by oscillations of the magnetic tip in a static inhomogeneous magnetic field $B_{\text{ring}}$ produced by the ring at the location of the magnetic tip,  is proportional to $\Delta f \propto (\partial^2B_{\text{ring}} / \partial x^2) m_{\text{tip}}$, where $m_{\text{tip}}$ is the magnetic moment of the tip~\cite{Hartmann1999}. 
Near $T_c$, the dynamical interactions of TAPS and the cantilever emerge and the jumps in the cantilever frequency, corresponding to discrete changes in the winding number, are replaced by sharp dips resulting from the dynamical interaction of TAPS with the cantilever.

\begin{figure}[h!]
\includegraphics{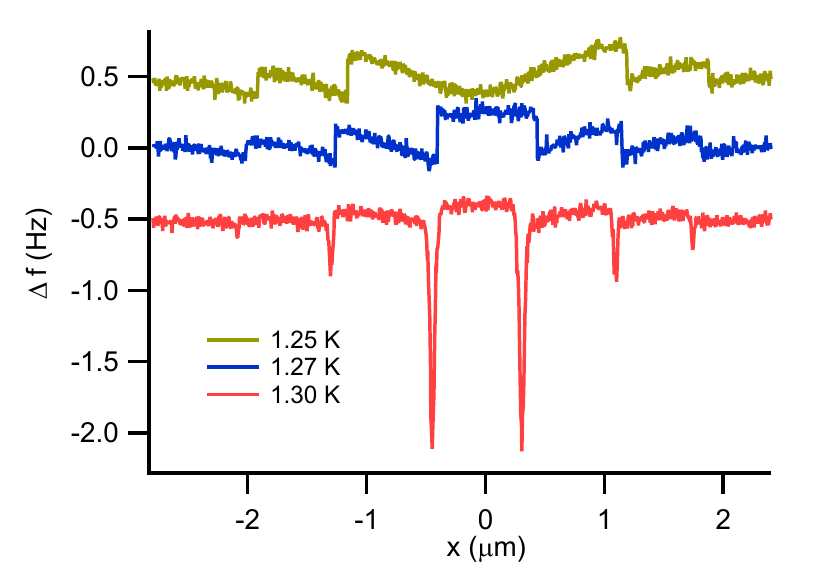}
\caption {Frequency shift data showing the transition from discrete fluxoid jumps to TAPS. The line scans were made along the diameter of the ring, at a tip-surface separation of 1.35~$\mu$m. A line scan taken above the superconducting transition was used to subtract the frequency background. The traces are offset for clarity.}
\label{fig:StepsDips}
\end{figure}

Figure ~\ref{fig:StepsDips} shows the transition from the jumps in frequency observed at lower temperatures to the dips in frequency caused by the dynamical interactions observed near $T_c$.
Line scans were obtained at several temperatures along the diameter of an aluminum ring with dimensions $R=0.95~\mu$m, wall width $w=100$~nm and critical temperature $T_c=1.32$~K.
As the data reveals, the dynamical effect produces much stronger frequency shifts than the one observed at low temperature.

\section{Estimate of the magnetic field distribution produced by the magnetic particle}\label{sec:TipField}

\begin{figure}[h!]
\includegraphics{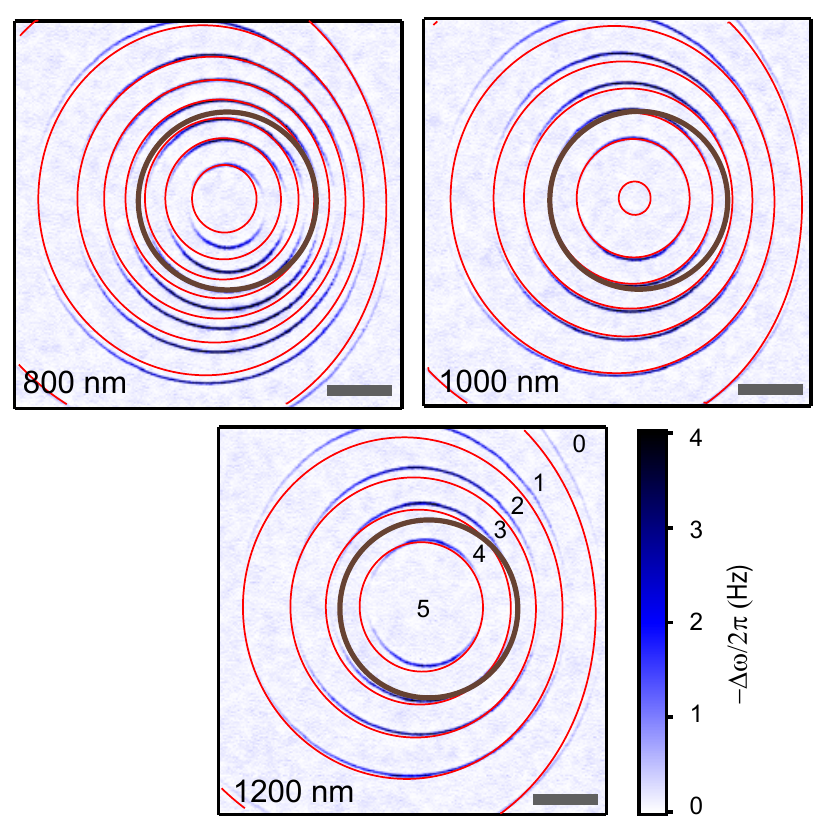}
\caption {Comparison of the fluxoid transitions calculated from the 3D model of the tip (red) to the experimental data. The \\textcolor[rgb]{0,0,1}{brown} circle represent the location of the ring. 
Calculated fluxoid states are labeled with the phase winding numbers on the image taken at tip-surface separation of $1200$~nm. 
All scale bars correspond to 1~$\mu$m.}
\label{fig:MatchedPatterns}
\end{figure}

In this appendix, we discuss the details for estimating the magnetic field profile produced by the magnetic particle for the measurements presented in Sec.~\ref{sec:PlainRing}.
From the SEM images of the tip, we modeled the geometry of the magnet particle as the sum of a cube having dimensions $(710\times 800\times 840)$~nm$^3$, and a pillar with dimensions $(290\times 290\times 1800)$~nm$^3$.
As a first approximation, we assume a uniformly magnetized tip. We determine the magnitude of the magnetic moment using cantilever torque magnetometry \cite{Stipe2001} to be $m_{\text{tip}}=7.2 \times 10^{-13}$~J/T. 
From this model of the tip, we calculate the positions of the contours that correspond to a half-integer number of flux quanta threading the ring, and we compare them to the frequency shift images taken for tip-surface separations of 800, 1000, and 1200~nm.
To achieve a good correspondence between the calculated and measured frequency shift contours, we vary the parameters of the model, e.g., the magnitude, orientation, and distribution of the magnetic moment.
The best agreement (Fig.~\ref{fig:MatchedPatterns}) is achieved by adjusting the magnetization of the pillar to be $0.25\times$ the magnetization of the cube, and by making the total magnetic moment of the particle to be $m_{\text{tip}}=4.7 \times 10^{-13}$~J/T.
The estimated magnetic moment corresponds to $\approx0.9$ of the maximum magnetic moment for this size particle, assuming the bulk magnetization of SmCo$_5$ of $M=0.84\times 10^{6}$~A/m \cite{Coey2002}.
The reduction of the magnetization in the pillar could  be caused by ion damage during the FIB micro-machining of the particle.

To account for the asymmetry observed in the frequency shift contours, we assume that the magnetic moment is tilted by $19^\circ$ in the $-y$~direction; the tip itself is tilted by $22^\circ$ sideways in the $+y$~direction.
The presence of multiple magnetic domains in the $\mathrm{SmCo_5}$ particle might explain the large tilt angles required to match the experimental data. 
The cross-sections of the field profiles are shown in Fig.~\ref{Ring}(e).

The fluxoid transition contours measured in the experiment were also matched by using an effective point dipole model of the tip. 
While it is possible to achieve very good agreement for data taken at a particular tip-surface height, the position of the point dipole must be varried for scans at different tip-surface separations. 
The estimates of the magnetic field with the effective point dipole model, when compared to those from the 3D model, give slightly broader distributions with $\sim 10\%$ lower peak magnetic fields under the tip. 
Hence, we suggest 10\% as an upper bound on the error for the estimation of the tip field.
The source of error is a combination of the complicated shape of the tip and the sparsity of the transition lines used in matching to the model. 
The precision of the calibration procedure could be improved in several ways: (i) using tips that have a simpler geometry, e.g., a bar having a uniform cross section; (ii) combining $N$ scans, each taken by applying a uniform magnetic field with magnitude $B_l=l\, \Phi_0/(N \pi R^2)$, where $l=\{0,1 ,\dots N-1\}$. This would increase the number of transition lines on the scan by $N$ times, and it would better constrain the tip model.

\section{Estimate of the coupling between the cantilever  and the supercurrent in the ring}\label{sec:TipRingCoupling}

The coupling between the cantilever and the supercurrent for a thin-wall ring $\kappa (\mathbf{r}_{tip})$ can be estimated by noticing that the mechanical work $ -\zeta \delta x$ needed to move the tip  by a distance $\delta x$ is equal to the magnetic energy $- I \delta \Phi$.
Hence, we find $\kappa(\mathbf{r}_{tip})=\Phi_0  (d\phi/dx)$. From Eq.(\ref{eq:alpha}), we obtain
\begin{equation}
\alpha=(d\phi/dx)^2\Phi_0/k,
\label{eq:alphaEstimate}
\end{equation}
which gives an estimate  $\alpha\simeq 38$~$\mathrm{A}^{-1}$.

Using Eq.~\eqref{eq:P_eq} and the relationship $I=-(1/\Phi_0)\partial F/\partial \phi$, we can express $\beta$ as
\begin{equation}\label{eq:IdPdf}
\beta=- \frac{\Phi_0\,\Delta I^2(\phi)}{4 k_B T} \cosh^{-2}\left\{ \frac{-[F_{n+1}(\phi)-F_{n}(\phi)]}{2 k_B T}\right\}. 
\end{equation}
At $\phi=n+1/2$  from \eqref{eq:Current} we obtain 
 \begin{equation}
\label{eq:DeltaI}
\Delta I= I_0 \left(1-\frac{\xi^2}{4 R^2}\right),
\end{equation}
and from Eqs.~\eqref{eq:IdPdf}
\begin{equation}
\beta=- \frac{\Phi_0}{4 k_B T} \left(1-\frac{\xi^2}{4 R^2}\right)^2 I_0^2. 
\label{eq:BetaCrossing}
\end{equation}

\begin{figure}[htbp]
\includegraphics{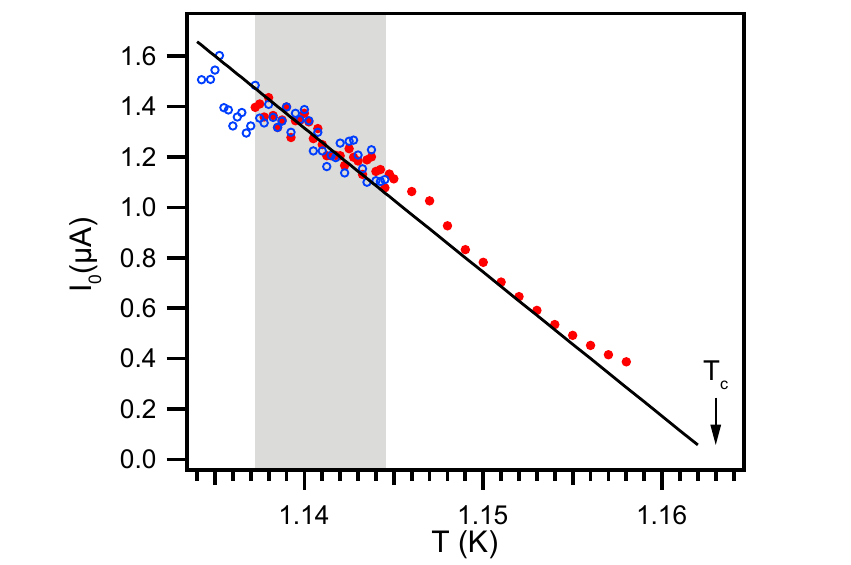}
\caption {Plot of $I_0(T)$ derived from $\Delta \gamma$ (empty circles) and $\Delta \omega$ (solid circles). Solid line corresponds to the fit of $I_0$ to the expected temperature dependence $I_0(T)\propto 1-T/T_c$. Shaded region marks the temperature range where $\nu_r (T)$ was extracted from data; points at other temperatures were calculated with extrapolated $\nu_r$ and were not used for the fit.
}
\label{fig:I0}
\end{figure}

Using Eqs.~(\ref{eq:Frequency}-\ref{eq:Damping}) together with the measured $\nu_r(T)$ (Eq.~\eqref{eq:ApproxRate}) and the estimate of $\alpha$ (Eq.~\eqref{eq:alphaEstimate}) we find $\beta(T)$ and hence $I_0(T)$ (Eq.~\eqref{eq:BetaCrossing}) from both $\Delta \omega$ and $\Delta \gamma$, as shown in Fig.~\ref{fig:I0}.
For points outside the temperature range where the relaxation rate was measured (marked by a gray band in Fig.~\ref{fig:I0}) $\nu_r(T)$ was extrapolated using Eq.~\eqref{eq:ApproxRate}.
By fitting $I_0(T)$ to the expected temperature dependence 
$I_0(T)\propto\lambda(T)^{-2}={\lambda (0)}^{-2} (1-t)$ we found that 
\begin{equation}
I_0(T)=66(1-t)\mu\text{A},
\label{eq:MeasuredCurrent}
\end{equation}
which corresponds to $\lambda(0)\simeq 164$~nm~(Fig.~\ref{fig:I0}).
The data points, for which $\nu_r$ was extrapolated were not used for the fit. 
The temperature dependences of $\Delta \omega(T)$ and $\Delta \gamma(T)$ calculated at $\phi=n+1/2$ using the SR model (Eqs.~(\ref{eq:Frequency}-\ref{eq:Damping}))  with $I_0(T)$ and $\nu_r(T)$ given by \eqref{eq:MeasuredCurrent} and \eqref{eq:ApproxRate} respectively, are plotted in Fig.~\ref{fig:StochasticResonance}(b) and describe the data well.

Comparison of the shape of the $\Delta \omega$ and $\Delta \gamma$ peaks, shown in Fig.~\ref{fig:StochasticResonance}(a), to the SR model requires $\beta(\phi)$ and $\nu_r(\phi)$.
Note, that for $\Delta \phi=\phi-(n+1/2)\ll 1$:
\begin{align}\label{eq:Shape}
&\Delta I(\phi)\simeq \Delta I\bigr|_{\phi=n+1/2},\\
\label{eq:ShapeEnergy}
&F_{n+1}(\phi)-F_{n}(\phi)\simeq -\Phi_0 \Delta I  \Delta \phi,\\
\label{eq:ShapeBeta}
&\beta(\phi)\simeq\beta\bigr|_{\phi=n+1/2}\, \cosh^{-2}\left\{ \frac{\Phi_0 \Delta I \Delta \phi}{2 k_B T}\right\},\\
&\nu_r(\phi)\simeq\nu_r\bigr|_{\phi=n+1/2}\, \cosh\left\{ \frac{\Phi_0 \Delta I \Delta \phi}{2 k_B T}\right\}.
\label{eq:ShapeRelaxRate}
\end{align} 
Here, Eq.~\eqref{eq:ShapeBeta} is found by combining \eqref{eq:IdPdf} and \eqref{eq:ShapeEnergy}. 
Approximation for $\nu_r(\phi)$ (Eq.~\eqref{eq:ShapeRelaxRate}) is obtained from Eqs.~\eqref{eq:RelaxRate}, \eqref{eq:PLrate} and \eqref{eq:ShapeEnergy} in the assumption that the saddle point free energy $F_{barrier}(\phi)$  that sets the phase slip barrier (see Fig.~\ref{fig:SRscheme}(a)) is flat around $\Delta \phi=0$: $\partial F_{barrier}/\partial \phi =0$.
Equations \eqref{eq:ShapeBeta}, \eqref{eq:ShapeRelaxRate}, \eqref{eq:DeltaI}, \eqref{eq:BetaCrossing} enable us to express $\beta(\phi,T)$ and $\nu_r(\phi, T)$ in terms of 
 $I_0(T)$ and $\nu_r(T)$ at $\Delta \phi=0$ which were determined earlier(\eqref{eq:MeasuredCurrent}, \eqref{eq:ApproxRate}).
The calculated $\Delta \omega(\phi)$ and $\Delta \gamma(\phi)$ curves are shown in Fig.~\ref{fig:StochasticResonance}(a) and are in a good agreement with the experimental data. 

In our analysis thus far, we have neglected the contribution to the flux from the self-inductance of the ring.
We estimate the self-inductance to be \cite{Grover1973}:
\begin{equation}
L\simeq\mu_0 R\left[\ln\left(\frac{8R}{w}\right)-\frac{1}{2}\right]= 6\,\text{pH}.
\end{equation}
From the signal strength we estimated that ${I_0(T)=(1-T/T_c)\times 66 \,\mu A}$. 
We can see that the circulating current has the largest value of $0.5\Delta I =0.6\,\mu A$ at $T\approx 1.14$~K.
The resulting correction to the applied flux from the self-inductance term is $1.7\times 10^{-3}\Phi_0$, which is sufficiently small so that we can neglect its contribution.

\begin{thebibliography}{36}%
\makeatletter
\providecommand \@ifxundefined [1]{%
 \@ifx{#1\undefined}
}%
\providecommand \@ifnum [1]{%
 \ifnum #1\expandafter \@firstoftwo
 \else \expandafter \@secondoftwo
 \fi
}%
\providecommand \@ifx [1]{%
 \ifx #1\expandafter \@firstoftwo
 \else \expandafter \@secondoftwo
 \fi
}%
\providecommand \natexlab [1]{#1}%
\providecommand \enquote  [1]{``#1''}%
\providecommand \bibnamefont  [1]{#1}%
\providecommand \bibfnamefont [1]{#1}%
\providecommand \citenamefont [1]{#1}%
\providecommand \href@noop [0]{\@secondoftwo}%
\providecommand \href [0]{\begingroup \@sanitize@url \@href}%
\providecommand \@href[1]{\@@startlink{#1}\@@href}%
\providecommand \@@href[1]{\endgroup#1\@@endlink}%
\providecommand \@sanitize@url [0]{\catcode `\\12\catcode `\$12\catcode
  `\&12\catcode `\#12\catcode `\^12\catcode `\_12\catcode `\%12\relax}%
\providecommand \@@startlink[1]{}%
\providecommand \@@endlink[0]{}%
\providecommand \url  [0]{\begingroup\@sanitize@url \@url }%
\providecommand \@url [1]{\endgroup\@href {#1}{\urlprefix }}%
\providecommand \urlprefix  [0]{URL }%
\providecommand \Eprint [0]{\href }%
\providecommand \doibase [0]{http://dx.doi.org/}%
\providecommand \selectlanguage [0]{\@gobble}%
\providecommand \bibinfo  [0]{\@secondoftwo}%
\providecommand \bibfield  [0]{\@secondoftwo}%
\providecommand \translation [1]{[#1]}%
\providecommand \BibitemOpen [0]{}%
\providecommand \bibitemStop [0]{}%
\providecommand \bibitemNoStop [0]{.\EOS\space}%
\providecommand \EOS [0]{\spacefactor3000\relax}%
\providecommand \BibitemShut  [1]{\csname bibitem#1\endcsname}%
\let\auto@bib@innerbib\@empty
\bibitem [{\citenamefont {Tinkham}(1996)}]{Tinkham1996}%
  \BibitemOpen
  \bibfield  {author} {\bibinfo {author} {\bibfnamefont {M.}~\bibnamefont
  {Tinkham}},\ }\href@noop {} {\emph {\bibinfo {title} {Introduction to
  superconductivity}}}\ (\bibinfo  {publisher} {Dover Publications}, \bibinfo
  {city} {Mineola},
  \bibinfo {year} {2004})\BibitemShut {NoStop}%
\bibitem [{\citenamefont {Parks}\ and\ \citenamefont
  {Little}(1964)}]{Parks1964}%
  \BibitemOpen
  \bibfield  {author} {\bibinfo {author} {\bibfnamefont {R.~D.}\ \bibnamefont
  {Parks}}\ and\ \bibinfo {author} {\bibfnamefont {W.~A.}\ \bibnamefont
  {Little}},\ }\href {\doibase 10.1103/PhysRev.133.A97} {\bibfield  {journal}
  {\bibinfo  {journal} {Phys. Rev.}\ }\textbf {\bibinfo {volume} {133}},\
  \bibinfo {pages} {A97} (\bibinfo {year} {1964})}\BibitemShut {NoStop}%
\bibitem [{\citenamefont {Morelle}\ \emph {et~al.}(2004)\citenamefont
  {Morelle}, \citenamefont {Golubovi\ifmmode~\acute{c}\else \'{c}\fi{}},\ and\
  \citenamefont {Moshchalkov}}]{Morelle2004}%
  \BibitemOpen
  \bibfield  {author} {\bibinfo {author} {\bibfnamefont {M.}~\bibnamefont
  {Morelle}}, \bibinfo {author} {\bibfnamefont {D.~S.}\ \bibnamefont
  {Golubovi\ifmmode~\acute{c}\else \'{c}\fi{}}}, \ and\ \bibinfo {author}
  {\bibfnamefont {V.~V.}\ \bibnamefont {Moshchalkov}},\ }\href {\doibase
  10.1103/PhysRevB.70.144528} {\bibfield  {journal} {\bibinfo  {journal} {Phys.
  Rev. B}\ }\textbf {\bibinfo {volume} {70}},\ \bibinfo {pages} {144528}
  (\bibinfo {year} {2004})}\BibitemShut {NoStop}%
\bibitem [{\citenamefont {Pedersen}\ \emph {et~al.}(2001)\citenamefont
  {Pedersen}, \citenamefont {Kofod}, \citenamefont {Hollingbery}, \citenamefont
  {S\o{}rensen},\ and\ \citenamefont {Lindelof}}]{Pedersen2001}%
  \BibitemOpen
  \bibfield  {author} {\bibinfo {author} {\bibfnamefont {S.}~\bibnamefont
  {Pedersen}}, \bibinfo {author} {\bibfnamefont {G.~R.}\ \bibnamefont {Kofod}},
  \bibinfo {author} {\bibfnamefont {J.~C.}\ \bibnamefont {Hollingbery}},
  \bibinfo {author} {\bibfnamefont {C.~B.}\ \bibnamefont {S\o{}rensen}}, \ and\
  \bibinfo {author} {\bibfnamefont {P.~E.}\ \bibnamefont {Lindelof}},\ }\href
  {\doibase 10.1103/PhysRevB.64.104522} {\bibfield  {journal} {\bibinfo
  {journal} {Phys. Rev. B}\ }\textbf {\bibinfo {volume} {64}},\ \bibinfo
  {pages} {104522} (\bibinfo {year} {2001})}\BibitemShut {NoStop}%
\bibitem [{\citenamefont {Vodolazov}\ \emph {et~al.}(2003)\citenamefont
  {Vodolazov}, \citenamefont {Peeters}, \citenamefont {Dubonos},\ and\
  \citenamefont {Geim}}]{Vodolazov2003}%
  \BibitemOpen
  \bibfield  {author} {\bibinfo {author} {\bibfnamefont {D.~Y.}\ \bibnamefont
  {Vodolazov}}, \bibinfo {author} {\bibfnamefont {F.~M.}\ \bibnamefont
  {Peeters}}, \bibinfo {author} {\bibfnamefont {S.~V.}\ \bibnamefont
  {Dubonos}}, \ and\ \bibinfo {author} {\bibfnamefont {A.~K.}\ \bibnamefont
  {Geim}},\ }\href {\doibase 10.1103/PhysRevB.67.054506} {\bibfield  {journal}
  {\bibinfo  {journal} {Phys. Rev. B}\ }\textbf {\bibinfo {volume} {67}},\
  \bibinfo {pages} {054506} (\bibinfo {year} {2003})}\BibitemShut {NoStop}%
\bibitem [{\citenamefont {Davidovi\ifmmode~\acute{c}\else \'{c}\fi{}}\ \emph
  {et~al.}(1996)\citenamefont {Davidovi\ifmmode~\acute{c}\else \'{c}\fi{}},
  \citenamefont {Kumar}, \citenamefont {Reich}, \citenamefont {Siegel},
  \citenamefont {Field}, \citenamefont {Tiberio}, \citenamefont {Hey},\ and\
  \citenamefont {Ploog}}]{Davidovic1996}%
  \BibitemOpen
  \bibfield  {author} {\bibinfo {author} {\bibfnamefont {D.}~\bibnamefont
  {Davidovi\ifmmode~\acute{c}\else \'{c}\fi{}}}, \bibinfo {author}
  {\bibfnamefont {S.}~\bibnamefont {Kumar}}, \bibinfo {author} {\bibfnamefont
  {D.~H.}\ \bibnamefont {Reich}}, \bibinfo {author} {\bibfnamefont
  {J.}~\bibnamefont {Siegel}}, \bibinfo {author} {\bibfnamefont {S.~B.}\
  \bibnamefont {Field}}, \bibinfo {author} {\bibfnamefont {R.~C.}\ \bibnamefont
  {Tiberio}}, \bibinfo {author} {\bibfnamefont {R.}~\bibnamefont {Hey}}, \ and\
  \bibinfo {author} {\bibfnamefont {K.}~\bibnamefont {Ploog}},\ }\href
  {\doibase 10.1103/PhysRevLett.76.815} {\bibfield  {journal} {\bibinfo
  {journal} {Phys. Rev. Lett.}\ }\textbf {\bibinfo {volume} {76}},\ \bibinfo
  {pages} {815} (\bibinfo {year} {1996})}\BibitemShut {NoStop}%
\bibitem [{\citenamefont {Silver}\ and\ \citenamefont
  {Zimmerman}(1967)}]{Silver1967}%
  \BibitemOpen
  \bibfield  {author} {\bibinfo {author} {\bibfnamefont {A.~H.}\ \bibnamefont
  {Silver}}\ and\ \bibinfo {author} {\bibfnamefont {J.~E.}\ \bibnamefont
  {Zimmerman}},\ }\href {\doibase 10.1103/PhysRev.157.317} {\bibfield
  {journal} {\bibinfo  {journal} {Phys. Rev.}\ }\textbf {\bibinfo {volume}
  {157}},\ \bibinfo {pages} {317} (\bibinfo {year} {1967})}\BibitemShut
  {NoStop}%
\bibitem [{\citenamefont {Lukens}\ and\ \citenamefont
  {Goodkind}(1968)}]{Lukens1968}%
  \BibitemOpen
  \bibfield  {author} {\bibinfo {author} {\bibfnamefont {J.~E.}\ \bibnamefont
  {Lukens}}\ and\ \bibinfo {author} {\bibfnamefont {J.~M.}\ \bibnamefont
  {Goodkind}},\ }\href {\doibase 10.1103/PhysRevLett.20.1363} {\bibfield
  {journal} {\bibinfo  {journal} {Phys. Rev. Lett.}\ }\textbf {\bibinfo
  {volume} {20}},\ \bibinfo {pages} {1363} (\bibinfo {year}
  {1968})}\BibitemShut {NoStop}%
\bibitem [{\citenamefont {Zhang}\ and\ \citenamefont
  {Price}(1997)}]{Zhang1997}%
  \BibitemOpen
  \bibfield  {author} {\bibinfo {author} {\bibfnamefont {X.}~\bibnamefont
  {Zhang}}\ and\ \bibinfo {author} {\bibfnamefont {J.~C.}\ \bibnamefont
  {Price}},\ }\href {\doibase 10.1103/PhysRevB.55.3128} {\bibfield  {journal}
  {\bibinfo  {journal} {Phys. Rev. B}\ }\textbf {\bibinfo {volume} {55}},\
  \bibinfo {pages} {3128} (\bibinfo {year} {1997})}\BibitemShut {NoStop}%
\bibitem [{\citenamefont {Kirtley}\ \emph {et~al.}(2003)\citenamefont
  {Kirtley}, \citenamefont {Tsuei}, \citenamefont {Kogan}, \citenamefont
  {Clem}, \citenamefont {Raffy},\ and\ \citenamefont {Li}}]{Kirtley2003}%
  \BibitemOpen
  \bibfield  {author} {\bibinfo {author} {\bibfnamefont {J.~R.}\ \bibnamefont
  {Kirtley}}, \bibinfo {author} {\bibfnamefont {C.~C.}\ \bibnamefont {Tsuei}},
  \bibinfo {author} {\bibfnamefont {V.~G.}\ \bibnamefont {Kogan}}, \bibinfo
  {author} {\bibfnamefont {J.~R.}\ \bibnamefont {Clem}}, \bibinfo {author}
  {\bibfnamefont {H.}~\bibnamefont {Raffy}}, \ and\ \bibinfo {author}
  {\bibfnamefont {Z.~Z.}\ \bibnamefont {Li}},\ }\href {\doibase
  10.1103/PhysRevB.68.214505} {\bibfield  {journal} {\bibinfo  {journal} {Phys.
  Rev. B}\ }\textbf {\bibinfo {volume} {68}},\ \bibinfo {pages} {214505}
  (\bibinfo {year} {2003})}\BibitemShut {NoStop}%
\bibitem [{\citenamefont {Bluhm}\ \emph {et~al.}(2006)\citenamefont {Bluhm},
  \citenamefont {Koshnick}, \citenamefont {Huber},\ and\ \citenamefont
  {Moler}}]{Bluhm2006}%
  \BibitemOpen
  \bibfield  {author} {\bibinfo {author} {\bibfnamefont {H.}~\bibnamefont
  {Bluhm}}, \bibinfo {author} {\bibfnamefont {N.~C.}\ \bibnamefont {Koshnick}},
  \bibinfo {author} {\bibfnamefont {M.~E.}\ \bibnamefont {Huber}}, \ and\
  \bibinfo {author} {\bibfnamefont {K.~A.}\ \bibnamefont {Moler}},\ }\href
  {\doibase 10.1103/PhysRevLett.97.237002} {\bibfield  {journal} {\bibinfo
  {journal} {Phys. Rev. Lett.}\ }\textbf {\bibinfo {volume} {97}},\ \bibinfo
  {pages} {237002} (\bibinfo {year} {2006})}\BibitemShut {NoStop}%
\bibitem [{\citenamefont {Koshnick}\ \emph {et~al.}(2007)\citenamefont
  {Koshnick}, \citenamefont {Bluhm}, \citenamefont {Huber},\ and\ \citenamefont
  {Moler}}]{Koshnick2007}%
  \BibitemOpen
  \bibfield  {author} {\bibinfo {author} {\bibfnamefont {N.~C.}\ \bibnamefont
  {Koshnick}}, \bibinfo {author} {\bibfnamefont {H.}~\bibnamefont {Bluhm}},
  \bibinfo {author} {\bibfnamefont {M.~E.}\ \bibnamefont {Huber}}, \ and\
  \bibinfo {author} {\bibfnamefont {K.~A.}\ \bibnamefont {Moler}},\ }\href
  {\doibase 10.1126/science.1148758} {\bibfield  {journal} {\bibinfo  {journal}
  {Science}\ }\textbf {\bibinfo {volume} {318}},\ \bibinfo {pages} {1440}
  (\bibinfo {year} {2007})}\BibitemShut {NoStop}%
\bibitem [{\citenamefont {Bert}\ \emph {et~al.}(2011)\citenamefont {Bert},
  \citenamefont {Koshnick}, \citenamefont {Bluhm},\ and\ \citenamefont
  {Moler}}]{Bert2011}%
  \BibitemOpen
  \bibfield  {author} {\bibinfo {author} {\bibfnamefont {J.~A.}\ \bibnamefont
  {Bert}}, \bibinfo {author} {\bibfnamefont {N.~C.}\ \bibnamefont {Koshnick}},
  \bibinfo {author} {\bibfnamefont {H.}~\bibnamefont {Bluhm}}, \ and\ \bibinfo
  {author} {\bibfnamefont {K.~A.}\ \bibnamefont {Moler}},\ }\href {\doibase
  10.1103/PhysRevB.84.134523} {\bibfield  {journal} {\bibinfo  {journal} {Phys.
  Rev. B}\ }\textbf {\bibinfo {volume} {84}},\ \bibinfo {pages} {134523}
  (\bibinfo {year} {2011})}\BibitemShut {NoStop}%
\bibitem [{\citenamefont {Bourgeois}\ \emph {et~al.}(2005)\citenamefont
  {Bourgeois}, \citenamefont {Skipetrov}, \citenamefont {Ong},\ and\
  \citenamefont {Chaussy}}]{Bourgeois2005}%
  \BibitemOpen
  \bibfield  {author} {\bibinfo {author} {\bibfnamefont {O.}~\bibnamefont
  {Bourgeois}}, \bibinfo {author} {\bibfnamefont {S.~E.}\ \bibnamefont
  {Skipetrov}}, \bibinfo {author} {\bibfnamefont {F.}~\bibnamefont {Ong}}, \
  and\ \bibinfo {author} {\bibfnamefont {J.}~\bibnamefont {Chaussy}},\ }\href
  {\doibase 10.1103/PhysRevLett.94.057007} {\bibfield  {journal} {\bibinfo
  {journal} {Phys. Rev. Lett.}\ }\textbf {\bibinfo {volume} {94}},\ \bibinfo
  {pages} {057007} (\bibinfo {year} {2005})}\BibitemShut {NoStop}%
\bibitem [{\citenamefont {Petkovi{\'c}}\ \emph {et~al.}(2016)\citenamefont
  {Petkovi{\'c}}, \citenamefont {Lollo}, \citenamefont {Glazman},\ and\
  \citenamefont {Harris}}]{Petkovic2016}%
  \BibitemOpen
  \bibfield  {author} {\bibinfo {author} {\bibfnamefont {I.}~\bibnamefont
  {Petkovi{\'c}}}, \bibinfo {author} {\bibfnamefont {A.}~\bibnamefont {Lollo}},
  \bibinfo {author} {\bibfnamefont {L.}~\bibnamefont {Glazman}}, \ and\
  \bibinfo {author} {\bibfnamefont {J.}~\bibnamefont {Harris}},\ }\href
  {\doibase 10.1038/ncomms13551} {\bibfield  {journal} {\bibinfo  {journal}
  {Nat. Commun.}\ }\textbf {\bibinfo {volume} {7}},\ \bibinfo {pages} {13551} (\bibinfo {year}
  {2016})}\BibitemShut {NoStop}%
\bibitem [{\citenamefont {Jang}\ \emph {et~al.}(2011)\citenamefont {Jang},
  \citenamefont {Ferguson}, \citenamefont {Vakaryuk}, \citenamefont {Budakian},
  \citenamefont {Chung}, \citenamefont {Goldbart},\ and\ \citenamefont
  {Maeno}}]{Jang2011}%
  \BibitemOpen
  \bibfield  {author} {\bibinfo {author} {\bibfnamefont {J.}~\bibnamefont
  {Jang}}, \bibinfo {author} {\bibfnamefont {D.~G.}\ \bibnamefont {Ferguson}},
  \bibinfo {author} {\bibfnamefont {V.}~\bibnamefont {Vakaryuk}}, \bibinfo
  {author} {\bibfnamefont {R.}~\bibnamefont {Budakian}}, \bibinfo {author}
  {\bibfnamefont {S.~B.}\ \bibnamefont {Chung}}, \bibinfo {author}
  {\bibfnamefont {P.~M.}\ \bibnamefont {Goldbart}}, \ and\ \bibinfo {author}
  {\bibfnamefont {Y.}~\bibnamefont {Maeno}},\ }\href {\doibase
  10.1126/science.1193839} {\bibfield  {journal} {\bibinfo  {journal}
  {Science}\ }\textbf {\bibinfo {volume} {331}},\ \bibinfo {pages} {186}
  (\bibinfo {year} {2011})}\BibitemShut {NoStop}%
\bibitem [{\citenamefont {Jackel}\ \emph {et~al.}(1974)\citenamefont {Jackel},
  \citenamefont {Webb}, \citenamefont {Lukens},\ and\ \citenamefont
  {Pei}}]{Jackel1974}%
  \BibitemOpen
  \bibfield  {author} {\bibinfo {author} {\bibfnamefont {L.~D.}\ \bibnamefont
  {Jackel}}, \bibinfo {author} {\bibfnamefont {W.~W.}\ \bibnamefont {Webb}},
  \bibinfo {author} {\bibfnamefont {J.~E.}\ \bibnamefont {Lukens}}, \ and\
  \bibinfo {author} {\bibfnamefont {S.~S.}\ \bibnamefont {Pei}},\ }\href
  {\doibase 10.1103/PhysRevB.9.115} {\bibfield  {journal} {\bibinfo  {journal}
  {Phys. Rev. B}\ }\textbf {\bibinfo {volume} {9}},\ \bibinfo {pages} {115}
  (\bibinfo {year} {1974})}\BibitemShut {NoStop}%
\bibitem [{\citenamefont {Baelus}\ \emph {et~al.}(2000)\citenamefont {Baelus},
  \citenamefont {Peeters},\ and\ \citenamefont {Schweigert}}]{Baelus2000}%
  \BibitemOpen
  \bibfield  {author} {\bibinfo {author} {\bibfnamefont {B.~J.}\ \bibnamefont
  {Baelus}}, \bibinfo {author} {\bibfnamefont {F.~M.}\ \bibnamefont {Peeters}},
  \ and\ \bibinfo {author} {\bibfnamefont {V.~A.}\ \bibnamefont {Schweigert}},\
  }\href {\doibase 10.1103/PhysRevB.61.9734} {\bibfield  {journal} {\bibinfo
  {journal} {Phys. Rev. B}\ }\textbf {\bibinfo {volume} {61}},\ \bibinfo
  {pages} {9734} (\bibinfo {year} {2000})}\BibitemShut {NoStop}%
\bibitem [{\citenamefont {Berger}(2003)}]{Berger2003}%
  \BibitemOpen
  \bibfield  {author} {\bibinfo {author} {\bibfnamefont {J.}~\bibnamefont
  {Berger}},\ }\href {\doibase 10.1103/PhysRevB.67.014531} {\bibfield
  {journal} {\bibinfo  {journal} {Phys. Rev. B}\ }\textbf {\bibinfo {volume}
  {67}},\ \bibinfo {pages} {014531} (\bibinfo {year} {2003})}\BibitemShut
  {NoStop}%
\bibitem [{\citenamefont {Kogan}\ \emph {et~al.}(2004)\citenamefont {Kogan},
  \citenamefont {Clem},\ and\ \citenamefont {Mints}}]{Kogan2004}%
  \BibitemOpen
  \bibfield  {author} {\bibinfo {author} {\bibfnamefont {V.~G.}\ \bibnamefont
  {Kogan}}, \bibinfo {author} {\bibfnamefont {J.~R.}\ \bibnamefont {Clem}}, \
  and\ \bibinfo {author} {\bibfnamefont {R.~G.}\ \bibnamefont {Mints}},\ }\href
  {\doibase 10.1103/PhysRevB.69.064516} {\bibfield  {journal} {\bibinfo
  {journal} {Phys. Rev. B}\ }\textbf {\bibinfo {volume} {69}},\ \bibinfo
  {pages} {064516} (\bibinfo {year} {2004})}\BibitemShut {NoStop}%
\bibitem [{\citenamefont {Vodolazov}\ and\ \citenamefont
  {Peeters}(2002)}]{Vodolazov2002b}%
  \BibitemOpen
  \bibfield  {author} {\bibinfo {author} {\bibfnamefont {D.~Y.}\ \bibnamefont
  {Vodolazov}}\ and\ \bibinfo {author} {\bibfnamefont {F.~M.}\ \bibnamefont
  {Peeters}},\ }\href {\doibase 10.1103/PhysRevB.66.054537} {\bibfield
  {journal} {\bibinfo  {journal} {Phys. Rev. B}\ }\textbf {\bibinfo {volume}
  {66}},\ \bibinfo {pages} {054537} (\bibinfo {year} {2002})}\BibitemShut
  {NoStop}%
\bibitem [{\citenamefont {Gammaitoni}\ \emph {et~al.}(1998)\citenamefont
  {Gammaitoni}, \citenamefont {H\"anggi}, \citenamefont {Jung},\ and\
  \citenamefont {Marchesoni}}]{Gammaintoni1998}%
  \BibitemOpen
  \bibfield  {author} {\bibinfo {author} {\bibfnamefont {L.}~\bibnamefont
  {Gammaitoni}}, \bibinfo {author} {\bibfnamefont {P.}~\bibnamefont
  {H\"anggi}}, \bibinfo {author} {\bibfnamefont {P.}~\bibnamefont {Jung}}, \
  and\ \bibinfo {author} {\bibfnamefont {F.}~\bibnamefont {Marchesoni}},\
  }\href {\doibase 10.1103/RevModPhys.70.223} {\bibfield  {journal} {\bibinfo
  {journal} {Rev. Mod. Phys.}\ }\textbf {\bibinfo {volume} {70}},\ \bibinfo
  {pages} {223} (\bibinfo {year} {1998})}\BibitemShut {NoStop}%
\bibitem [{\citenamefont {Woodside}\ and\ \citenamefont
  {McEuen}(2002)}]{Woodside2002}%
  \BibitemOpen
  \bibfield  {author} {\bibinfo {author} {\bibfnamefont {M.~T.}\ \bibnamefont
  {Woodside}}\ and\ \bibinfo {author} {\bibfnamefont {P.~L.}\ \bibnamefont
  {McEuen}},\ }\href {\doibase 10.1126/science.1069923} {\bibfield  {journal}
  {\bibinfo  {journal} {Science}\ }\textbf {\bibinfo {volume} {296}},\ \bibinfo
  {pages} {1098} (\bibinfo {year} {2002})}\BibitemShut {NoStop}%
\bibitem [{\citenamefont {Zhu}\ \emph {et~al.}(2005)\citenamefont {Zhu},
  \citenamefont {Brink},\ and\ \citenamefont {McEuen}}]{Zhu2005}%
  \BibitemOpen
  \bibfield  {author} {\bibinfo {author} {\bibfnamefont {J.}~\bibnamefont
  {Zhu}}, \bibinfo {author} {\bibfnamefont {M.}~\bibnamefont {Brink}}, \ and\
  \bibinfo {author} {\bibfnamefont {P.~L.}\ \bibnamefont {McEuen}},\ }\href
  {\doibase 10.1063/1.2139623} {\bibfield  {journal} {\bibinfo  {journal}
  {Applied Physics Letters}\ }\textbf {\bibinfo {volume} {87}},\ \bibinfo {eid}
  {242102} (\bibinfo {year} {2005}),\ 10.1063/1.2139623}\BibitemShut {NoStop}%
\bibitem [{\citenamefont {Zhu}\ \emph {et~al.}(2008)\citenamefont {Zhu},
  \citenamefont {Brink},\ and\ \citenamefont {McEuen}}]{Zhu2008}%
  \BibitemOpen
  \bibfield  {author} {\bibinfo {author} {\bibfnamefont {J.}~\bibnamefont
  {Zhu}}, \bibinfo {author} {\bibfnamefont {M.}~\bibnamefont {Brink}}, \ and\
  \bibinfo {author} {\bibfnamefont {P.~L.}\ \bibnamefont {McEuen}},\ }\href
  {\doibase 10.1021/nl801295y} {\bibfield  {journal} {\bibinfo  {journal} {Nano
  Letters}\ }\textbf {\bibinfo {volume} {8}},\ \bibinfo {pages} {2399}
  (\bibinfo {year} {2008})},\ \bibinfo {note} {pMID: 18578552}\BibitemShut
  {NoStop}%
\bibitem [{\citenamefont {Bennett}\ \emph {et~al.}(2010)\citenamefont
  {Bennett}, \citenamefont {Cockins}, \citenamefont {Miyahara}, \citenamefont
  {Gr\"utter},\ and\ \citenamefont {Clerk}}]{Bennett2010}%
  \BibitemOpen
  \bibfield  {author} {\bibinfo {author} {\bibfnamefont {S.~D.}\ \bibnamefont
  {Bennett}}, \bibinfo {author} {\bibfnamefont {L.}~\bibnamefont {Cockins}},
  \bibinfo {author} {\bibfnamefont {Y.}~\bibnamefont {Miyahara}}, \bibinfo
  {author} {\bibfnamefont {P.}~\bibnamefont {Gr\"utter}}, \ and\ \bibinfo
  {author} {\bibfnamefont {A.~A.}\ \bibnamefont {Clerk}},\ }\href {\doibase
  10.1103/PhysRevLett.104.017203} {\bibfield  {journal} {\bibinfo  {journal}
  {Phys. Rev. Lett.}\ }\textbf {\bibinfo {volume} {104}},\ \bibinfo {pages}
  {017203} (\bibinfo {year} {2010})}\BibitemShut {NoStop}%
\bibitem [{\citenamefont {Cockins}\ \emph {et~al.}(2010)\citenamefont
  {Cockins}, \citenamefont {Miyahara}, \citenamefont {Bennett}, \citenamefont
  {Clerk}, \citenamefont {Studenikin}, \citenamefont {Poole}, \citenamefont
  {Sachrajda},\ and\ \citenamefont {Grutter}}]{Cockins2010}%
  \BibitemOpen
  \bibfield  {author} {\bibinfo {author} {\bibfnamefont {L.}~\bibnamefont
  {Cockins}}, \bibinfo {author} {\bibfnamefont {Y.}~\bibnamefont {Miyahara}},
  \bibinfo {author} {\bibfnamefont {S.~D.}\ \bibnamefont {Bennett}}, \bibinfo
  {author} {\bibfnamefont {A.~A.}\ \bibnamefont {Clerk}}, \bibinfo {author}
  {\bibfnamefont {S.}~\bibnamefont {Studenikin}}, \bibinfo {author}
  {\bibfnamefont {P.}~\bibnamefont {Poole}}, \bibinfo {author} {\bibfnamefont
  {A.}~\bibnamefont {Sachrajda}}, \ and\ \bibinfo {author} {\bibfnamefont
  {P.}~\bibnamefont {Grutter}},\ }\href {\doibase 10.1073/pnas.0912716107}
  {\bibfield  {journal} {\bibinfo  {journal} {Proceedings of the National
  Academy of Sciences}\ }\textbf {\bibinfo {volume} {107}},\ \bibinfo {pages}
  {9496} (\bibinfo {year} {2010})}\BibitemShut {NoStop}%
\bibitem [{\citenamefont {Roy-Gobeil}\ \emph {et~al.}(2015)\citenamefont
  {Roy-Gobeil}, \citenamefont {Miyahara},\ and\ \citenamefont
  {Grutter}}]{RoyGobeil2015}%
  \BibitemOpen
  \bibfield  {author} {\bibinfo {author} {\bibfnamefont {A.}~\bibnamefont
  {Roy-Gobeil}}, \bibinfo {author} {\bibfnamefont {Y.}~\bibnamefont
  {Miyahara}}, \ and\ \bibinfo {author} {\bibfnamefont {P.}~\bibnamefont
  {Grutter}},\ }\href {\doibase 10.1021/nl504468a} {\bibfield  {journal}
  {\bibinfo  {journal} {Nano Letters}\ }\textbf {\bibinfo {volume} {15}},\
  \bibinfo {pages} {2324} (\bibinfo {year} {2015})}\BibitemShut {NoStop}%
\bibitem [{\citenamefont {Langer}\ and\ \citenamefont
  {Ambegaokar}(1967)}]{Langer1967}%
  \BibitemOpen
  \bibfield  {author} {\bibinfo {author} {\bibfnamefont {J.~S.}\ \bibnamefont
  {Langer}}\ and\ \bibinfo {author} {\bibfnamefont {V.}~\bibnamefont
  {Ambegaokar}},\ }\href {\doibase 10.1103/PhysRev.164.498} {\bibfield
  {journal} {\bibinfo  {journal} {Phys. Rev.}\ }\textbf {\bibinfo {volume}
  {164}},\ \bibinfo {pages} {498} (\bibinfo {year} {1967})}\BibitemShut
  {NoStop}%
\bibitem [{\citenamefont {McCumber}\ and\ \citenamefont
  {Halperin}(1970)}]{McCumber1970}%
  \BibitemOpen
  \bibfield  {author} {\bibinfo {author} {\bibfnamefont {D.~E.}\ \bibnamefont
  {McCumber}}\ and\ \bibinfo {author} {\bibfnamefont {B.~I.}\ \bibnamefont
  {Halperin}},\ }\href {\doibase 10.1103/PhysRevB.1.1054} {\bibfield  {journal}
  {\bibinfo  {journal} {Phys. Rev. B}\ }\textbf {\bibinfo {volume} {1}},\
  \bibinfo {pages} {1054} (\bibinfo {year} {1970})}\BibitemShut {NoStop}%
\bibitem [{\citenamefont {Albrecht}\ \emph {et~al.}(1991)\citenamefont
  {Albrecht}, \citenamefont {Grütter}, \citenamefont {Horne},\ and\
  \citenamefont {Rugar}}]{Albrecht1991}%
  \BibitemOpen
  \bibfield  {author} {\bibinfo {author} {\bibfnamefont {T.~R.}\ \bibnamefont
  {Albrecht}}, \bibinfo {author} {\bibfnamefont {P.}~\bibnamefont {Grütter}},
  \bibinfo {author} {\bibfnamefont {D.}~\bibnamefont {Horne}}, \ and\ \bibinfo
  {author} {\bibfnamefont {D.}~\bibnamefont {Rugar}},\ }\href {\doibase
  http://dx.doi.org/10.1063/1.347347} {\bibfield  {journal} {\bibinfo
  {journal} {Journal of Applied Physics}\ }\textbf {\bibinfo {volume} {69}},\
  \bibinfo {pages} {668} (\bibinfo {year} {1991})}\BibitemShut {NoStop}%
\bibitem [{\citenamefont {Little}(1967)}]{Little1967}%
  \BibitemOpen
  \bibfield  {author} {\bibinfo {author} {\bibfnamefont {W.~A.}\ \bibnamefont
  {Little}},\ }\href {\doibase 10.1103/PhysRev.156.396} {\bibfield  {journal}
  {\bibinfo  {journal} {Phys. Rev.}\ }\textbf {\bibinfo {volume} {156}},\
  \bibinfo {pages} {396} (\bibinfo {year} {1967})}\BibitemShut {NoStop}%
\bibitem [{\citenamefont {Stipe}\ \emph {et~al.}(2001)\citenamefont {Stipe},
  \citenamefont {Mamin}, \citenamefont {Stowe}, \citenamefont {Kenny},\ and\
  \citenamefont {Rugar}}]{Stipe2001}%
  \BibitemOpen
  \bibfield  {author} {\bibinfo {author} {\bibfnamefont {B.~C.}\ \bibnamefont
  {Stipe}}, \bibinfo {author} {\bibfnamefont {H.~J.}\ \bibnamefont {Mamin}},
  \bibinfo {author} {\bibfnamefont {T.~D.}\ \bibnamefont {Stowe}}, \bibinfo
  {author} {\bibfnamefont {T.~W.}\ \bibnamefont {Kenny}}, \ and\ \bibinfo
  {author} {\bibfnamefont {D.}~\bibnamefont {Rugar}},\ }\href {\doibase
  10.1103/PhysRevLett.86.2874} {\bibfield  {journal} {\bibinfo  {journal}
  {Phys. Rev. Lett.}\ }\textbf {\bibinfo {volume} {86}},\ \bibinfo {pages}
  {2874} (\bibinfo {year} {2001})}\BibitemShut {NoStop}%
\bibitem [{\citenamefont {Hartmann}(1999)}]{Hartmann1999}%
  \BibitemOpen
  \bibfield  {author} {\bibinfo {author} {\bibfnamefont {U.}~\bibnamefont
  {Hartmann}},\ }\href {\doibase 10.1146/annurev.matsci.29.1.53} {\bibfield
  {journal} {\bibinfo  {journal} {Annual Review of Materials Science}\ }\textbf
  {\bibinfo {volume} {29}},\ \bibinfo {pages} {53} (\bibinfo {year}
  {1999})}\BibitemShut {NoStop}%
\bibitem [{\citenamefont {Coey}(2002)}]{Coey2002}%
  \BibitemOpen
  \bibfield  {author} {\bibinfo {author} {\bibfnamefont {J.}~\bibnamefont
  {Coey}},\ }\href@noop {} {\bibfield  {journal} {\bibinfo  {journal} {Journal
  of Magnetism and Magnetic Materials}\ }\textbf {\bibinfo {volume} {248}},\
  \bibinfo {pages} {441} (\bibinfo {year} {2002})}\BibitemShut {NoStop}%
\bibitem [{\citenamefont {Grover}(1973)}]{Grover1973}%
  \BibitemOpen
  \bibfield  {author} {\bibinfo {author} {\bibfnamefont {F.~W.}\ \bibnamefont
  {Grover}},\ }\href@noop {} {\emph {\bibinfo {title} {Inductance
  calculations}}}\ (\bibinfo  {publisher} {Instrument Society of America,
  Research Triangle Park},\ \bibinfo {year} {1973})\ p.\ \bibinfo {pages}
  {143}\BibitemShut {NoStop}%
\end{thebibliography}
%

\end{document}